\documentclass[aps,pra,reprint,superscriptaddress]{revtex4-1}
\pdfoutput=1
\usepackage{graphicx}
\usepackage{amsmath,amsfonts,amssymb}
\usepackage{siunitx}

\begin{document}

\title{RF-induced heating dynamics of non-crystallized trapped ions}

\author{Martin W. van Mourik}
\author{Pavel Hrmo}
\author{Lukas Gerster}
\author{Benjamin Wilhelm}
\affiliation{Institut f\"ur Experimentalphysik, Universit\"at Innsbruck, Technikerstraße 25/4, 6020 Innsbruck, Austria}
\author{Rainer Blatt}
\affiliation{Institut f\"ur Experimentalphysik, Universit\"at Innsbruck, Technikerstraße 25/4, 6020 Innsbruck, Austria}
\affiliation{Institut f\"ur Quantenoptik und Quanteninformation, \"Osterreichische Akademie der Wissenschaften, Technikerstraße 21a, 6020 Innsbruck, Austria}
\author{Philipp Schindler}
\affiliation{Institut f\"ur Experimentalphysik, Universit\"at Innsbruck, Technikerstraße 25/4, 6020 Innsbruck, Austria}
\author{Thomas Monz}
\affiliation{Institut f\"ur Experimentalphysik, Universit\"at Innsbruck, Technikerstraße 25/4, 6020 Innsbruck, Austria}
\affiliation{AQT, Technikerstraße 17, 6020 Innsbruck, Austria}

\begin{abstract}  
We investigate the energy dynamics of non-crystallized (melted) ions, confined in a Paul trap. The non-periodic Coulomb interaction experienced by melted ions forms a medium for non-conservative energy transfer from the radio-frequency (rf) field to the ions, a process known as rf heating. We study rf heating by analyzing numerical simulations of non-crystallized ion motion in Paul trap potentials, in which the energy of the ions' secular motion changes at discrete intervals, corresponding to ion-ion collisions. The analysis of these collisions is used as a basis to derive a simplified model of rf heating energy dynamics, from which we conclude that the rf heating rate is predominantly dependent on the rf field strength. We confirm the predictability of the model experimentally: Two trapped $^{40}$Ca$^{+}$ ions are deterministically driven to melt, and their fluorescence rate is used to infer the ions' energy. From simulation and experimental results, we generalize which experimental parameters are required for efficient recrystallization of melted trapped ions.
\end{abstract}

\maketitle

\section{Introduction}
\label{sec:Introduction}
Ions confined in radio-frequency (rf) Paul traps have enabled research in many fields of physics \cite{Leibfried_RevModPhys_2003}, such as fundamental light-matter interactions \cite{Araneda2018,Monroe1996}, frequency measurements and metrology \cite{Poli2013}, mass spectrometry \cite{Douglas05}, quantum computation \cite{Bruzewicz2019,Debnath2016,Bermudez2017,Monz2015}, and quantum simulation \cite{Schindler2013,Blatt2012a,Zhang2017}. The vast majority of modern trapped-ion experiments utilize a so-called ``ion crystal'', a regular spatial structure of multiple ions. Such experiments rely on the fact that these crystals contain well-localized separated particles that share common motional modes due to their Coulomb interaction. For example, in trapped ion-based quantum computation, an ion crystal represents a qubit register, and their common motional modes are the data-bus that mediates entanglement \cite{Steane_ApplPhysB_1997,James_PhysRevA_2001}.

A prominent event that disturbs the crystal structure is a collision with a particle from the residual background gas \cite{Hankin_PhysRevA_2019}. Such a collision can transfer enough energy to ions such that the crystal structure is destroyed. The ions undergo a transition described as melting, to a phase colloquially named an ion cloud \cite{Blumel_Nature_1988,Blumel_PRA_1989}, and are no longer suitable to be used as qubits for quantum computation.

Melted ions are subjected to a change in energy that is not present in the crystal phase: Energy can be transferred to the ions from the rf field from the trapping electrodes, leading to an overall increase in the ions' average kinetic energy. This process, known as \emph{rf heating}, occurs when ions experience both non-periodic Coulomb forces and forces from the trap's oscillating rf field. Rf heating has previously been studied in the context of interactions of ions with ultra-cold buffer gasses \cite{Cetina_PhysRevLett_2012,Nguyen_PhysRevA_2012}. Rf heating due to ion-ion interactions is a dominant source of energy gain in ion clouds, but has not been studied in detail, despite melting being a common occurrence in ion trap experiments. 

The performance of trapped-ion experiments benefits from efficiently returning an ion cloud into the crystal state, a process known as \emph{recrystallization}. While laser cooling techniques can be employed to remove energy from the ions, the opposing increase in energy due to rf heating hinders or even prevents recrystallization. 

In this work, we study the dynamics of rf heating in ion clouds. As the motion of melted ions in an rf field is chaotic \cite{Blumel_PRA_1989,Gottwald_Chaos_2016}, it is inconceivable to attain generalized analytic descriptions of the ion motion. However, we can numerically analyze dynamics of melted ions with multiple initial conditions, from which we derive simplified models that provide quantitative approximations of the effects of rf heating. This approach allows us to determine laser cooling parameters required for overcoming rf heating, such that ions recrystallize.

\begin{figure}
    \centering
    \includegraphics[width=\linewidth]{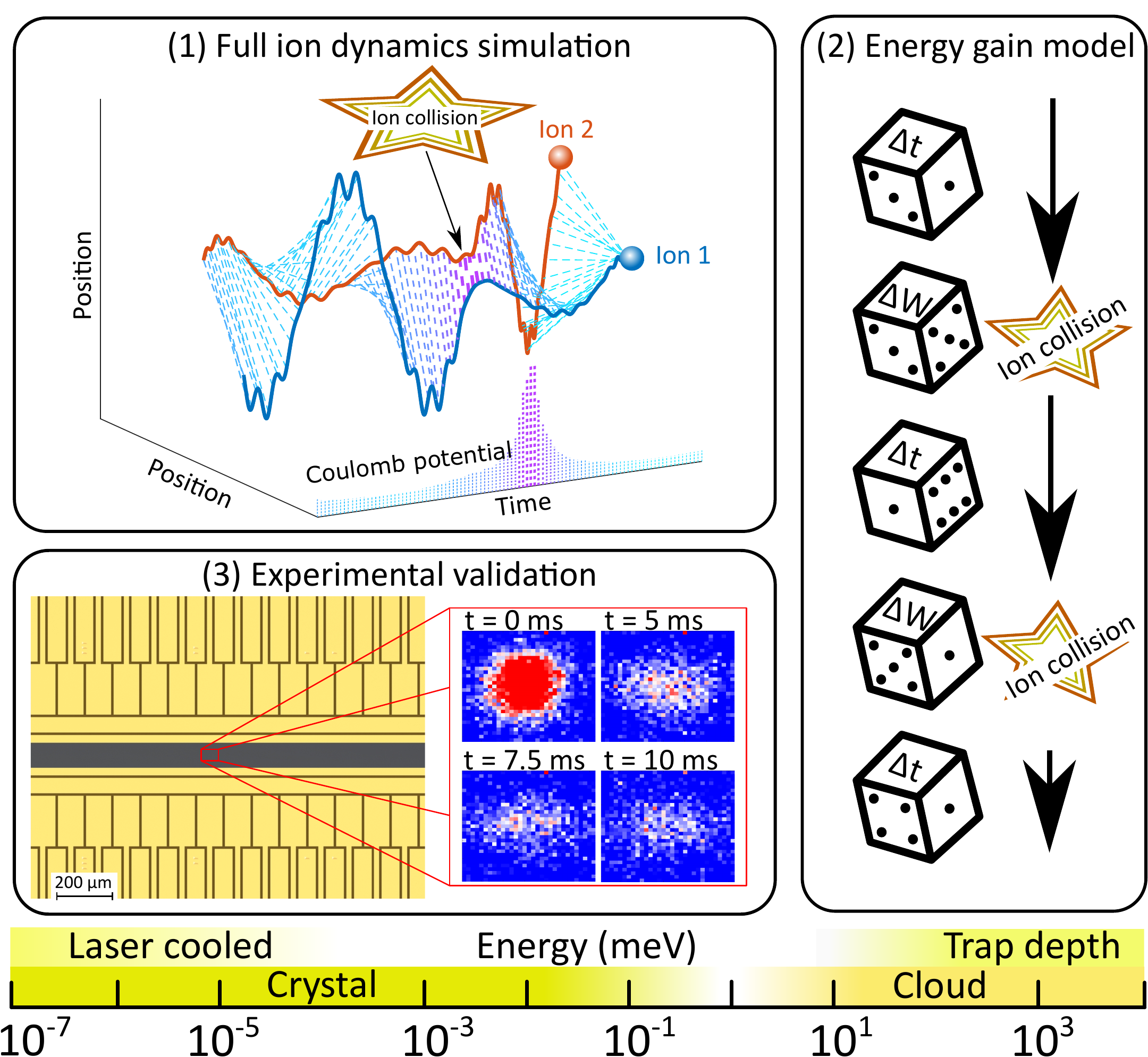}
    \caption{Schematic outline: (1) We simulate the full 3D dynamics of ions in a time-dependent field allowing us to numerically analyze the processes that lead to energy change. (2) Our analysis is used as a basis for a simplified model that describes the rf heating rate. (3) We experimentally create controlled melting events, from which we estimate the ion cloud energy. For reference, typical energy scales are displayed for relevant regimes.}
    \label{fig:paper_overview}
\end{figure}

This manuscript is structured as follows: in Section \ref{sec:heating_overview} we provide a general description of the process of rf heating. Subsequent sections (\ref{sec:full_ion_dynamics_simulations} - \ref{sec:experimental}) detail our investigation of rf heating in three steps, as schematically depicted in Figure \ref{fig:paper_overview}: \textbf{1)} We use numerical simulations that track the motion of ions in a Paul trap to investigate their dynamics under the influence of an rf field and Coulomb interaction. From these simulations we surmise that changes in energy due to rf heating occurs at discrete moments in time, corresponding to small ion-ion distances that lead to a large Coulomb repulsion. \textbf{2)} We derive analytical expressions that approximate these energy changes and the intervals at which they occur. These expressions are the basis for a simplified simulation of ion cloud energy dynamics, that avoids the computational overhead involved in tracking the motion of all ions in an rf field. We use the simplified simulation to investigate how various trap parameters affect rf heating. Additionally, we investigate which Doppler cooling parameters can overcome rf heating to recrystallize ions. \textbf{3)} We experimentally validate our simulated results by a controlled melting of ion crystals and estimate the ions' energy change by monitoring changes in the cloud's fluorescence.

\section{Heating in time-dependent potentials: rf heating}
\label{sec:heating_overview}
In this section we give a general overview of the process of rf heating. We provide a qualitative description of the physical process involved in energy exchange between the rf field and an ion's kinetic energy.

It is useful to describe the motion of trapped ions in two distinct timescales: rf-motion (or \emph{micromotion}), which describes the oscillation synchronous with the rf field, and \emph{secular motion}, which describes the motion in a static harmonic potential, known as the \emph{pseudopotential} \cite{Drewsen_PhysRevA_2000}. The total energy of the ions can be separated into the energies corresponding to these two timescales:
\begin{equation}
    \label{eq:energy_expansion}
    E_{tot} = E_{\mathrm{rf}}+E_{\mathrm{sec}}
\end{equation}
The total system energy $E_{tot}$ is the sum of contributions from the time-varying and static electric potential, the Coulomb interaction between ions, and the ions' kinetic energy.

In the crystal phase, the secular and rf components of energy do not couple with each other. The secular energy $E_{\mathrm{sec}}$ is then conserved or well-controlled, despite a time-dependent rf energy. In this regime, multiple co-trapped ions experience only small excursions from their respective trapping locations. Excursions are considered small if the ions' deviations from their trapping locations are much smaller than the inter-ion separation in an ion crystal. The motion of the ions can then be expanded into normal modes of motion, with distinct frequencies and ideally negligible coupling. 
The secular modes of motion and micromotion have unique frequencies in separate timescales, and thus remain separated. In fact, when assessing the ions' motion, micromotion is often neglected. In this secular approximation, the ions motion and energy are treated as if solely in an effective static harmonic potential, the pseudopotential.

In practice, undesired external influences can alter the ions' secular energy. For example, particles from the residual background gas can collide with an ion. Such a collision can transfer enough energy to ions such that their range of motion exceeds the crystal's inter-ion separation. Ions then undergo a transition to a melted state, an ion cloud. The motion of ions in a cloud is fundamentally different than in the crystal state. The average position of ions is fully governed by trap potentials, and not by Coulomb interaction. Furthermore, ions experience aperiodic motion due to irregular ion-ion Coulomb interactions. The frequency spectra corresponding to the secular motion and micromotion broaden and overlap due to this aperiodic motion. This allows energy from micromotion $E_{\mathrm{rf}}$ to be transferred to the secular motion $E_{\mathrm{sec}}$. Therefore, unlike for an ion crystal, micromotion cannot be neglected when considering motion of ions in a cloud, and the pseudopotential approximation is no longer valid.

\begin{figure}
    \includegraphics[width=\linewidth]{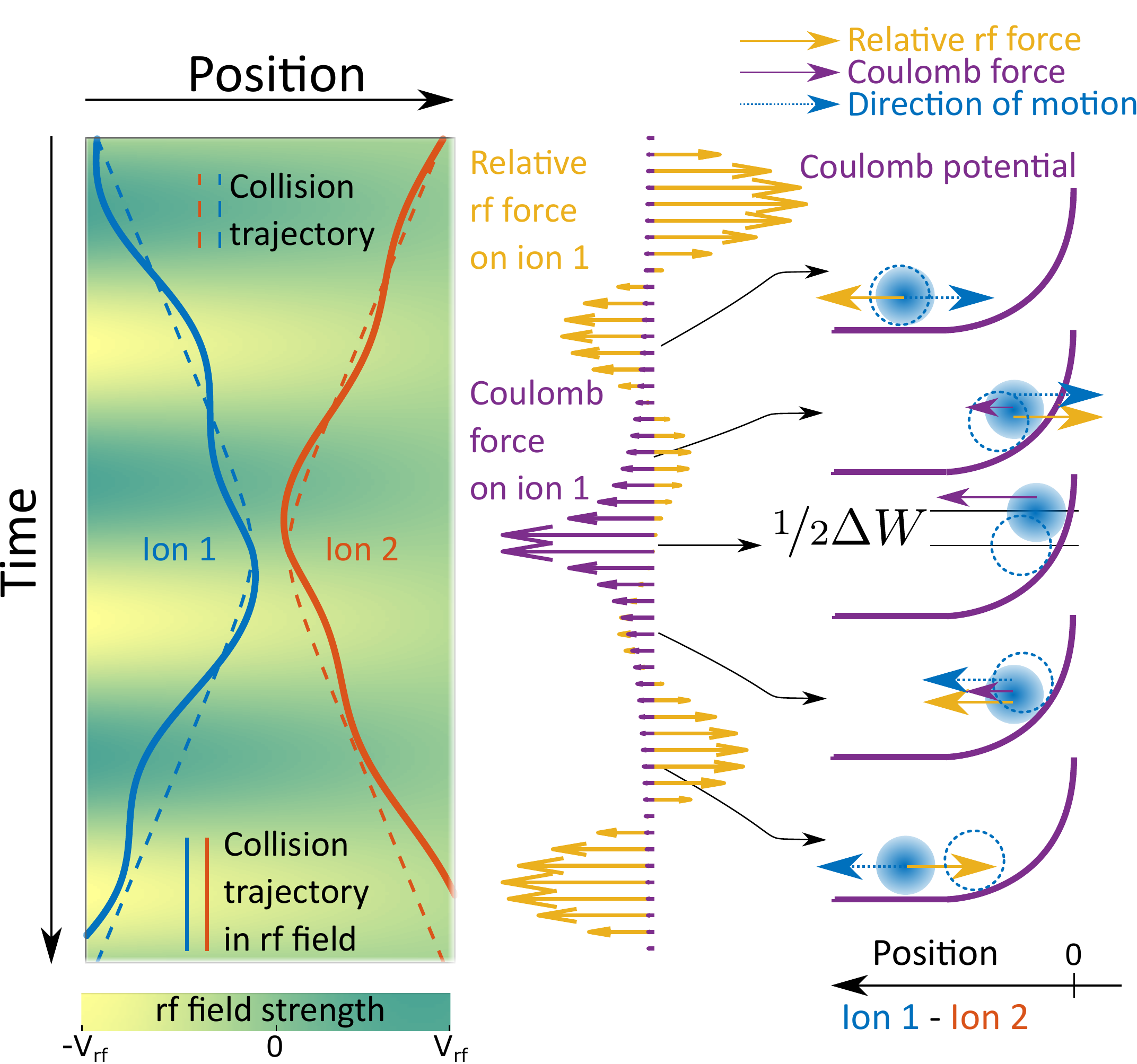}
    \caption{Schematic of energy changes due to Coulomb interaction in an rf field. While following a trajectory governed by Coulomb repulsion, ions additionally experience an unequal force from the rf field, due to its position dependence. In this example, the rf phase is tuned such that during the moments of increased Coulomb interaction, the relative rf field is aligned with the direction of motion, both during the approach and withdrawal in the ions motion. This leads to an increase of $\Delta W$ in motional energy after the collision. On the right, the filled circle represents the position of ion 1 as increases and decreases in the Coulomb potential, whereas the dotted circle is the position if no rf is applied.}
    \label{fig:collision_schematic}
\end{figure}

The energy transfer process from the rf-driven micromotion to the secular motion is known as rf heating, and is schematically outlined in one dimension in Figure \ref{fig:collision_schematic}. In a static potential, two ions would approach each other, experiencing opposing Coulomb forces, and repel, as denoted by the dashed lines. In an oscillating potential, ions deviate approximately sinusoidally (solid lines) from this path. Since the strength of the rf field is dependent on the ions' positions within the trap, the two ions experience different forces from the oscillating rf potential. In the example in Figure \ref{fig:collision_schematic}, directly before the moment of closest proximity, the difference in rf force reduces the distance between the ions compared to the static potential. Ions therefore have more Coulomb energy at small distances than they would have in a static potential. As the ions begin to repel, the rf field has switched sign, and now the difference in rf force aids in separating the ions. The extra relative velocity that the ions have gained by this time-varying force results in a gain in total energy in the ions' secular motion. The rf field has thus added energy (denoted as $\Delta W$ in Figure \ref{fig:collision_schematic}) to $E_{\mathrm{sec}}$. This process would remove energy from $E_{\mathrm{sec}}$ if the phase of the rf field had been shifted by $\pi$.

The schematic in Figure \ref{fig:collision_schematic} provides a qualitative description of the mechanism of energy transfer. In practice, such ``head on'' encounters do not occur in three dimensions, and the rf phase will generally not line up with the Coulomb force as schematically presented. 
In the following section, we simulate trapped ion trajectories in a time-varying potential and assess rf heating in three dimensions. 

\section{Full ion dynamics simulations}
\label{sec:full_ion_dynamics_simulations}
In this section, we investigate rf heating by analyzing simulations of particle dynamics of ion clouds. We numerically integrate the classical motion of charged particles by calculating forces given by a static (DC) field, a dynamic (rf) field, and Coulomb interactions. We refer to these simulations as ``full'', to contrast them with simplified simulations later in this work.

In our simulation, we track the dynamics of two trapped $^{40}$Ca$^{+}$ ions. We use trapping parameters that match typical experimental values \cite{Brandl2016}, with motional frequencies of $\lbrace\omega_x,\omega_y,\omega_z\rbrace = 2\pi \lbrace3.1,3.4,1.1\rbrace$ MHz, where our coordinate system is chosen such that $z$ corresponds to the direction with no rf potential (axial) and $x$ and $y$ the two radial directions. The fields in the simulation are time-dependent quadrupole potentials. At the start of a simulation, ions are placed in their crystallized equilibrium positions. One ion is given an initial velocity in a random direction, mimicking a collision with a background gas particle. An initial kinetic energy of 1.4 meV is chosen, as it is marginally more than the required energy to melt the crystal \cite{Prestage_PhysRevLett_1991}. This energy range corresponds to more than $10^5$ motional quanta ($\gtrapprox$ 16 K), so a classic evaluation of the equations of motion is justified.
Laser cooling is not included in these simulations. 

To investigate rf heating in these simulations, we determine the system's energy: In an ion cloud, each ion (with index $i$) has an energy given by the sum of its kinetic energy $V_{\mathrm{kin}}$ and its potential energy due to the trap's applied static and rf fields, $V_{\mathrm{DC}}$ and $V_{\mathrm{rf}}$. Additionally, ions have energy due to the Coulomb interaction potential $V_{\mathrm{Coul}}$ between particles. The total energy $E_{\mathrm{tot}}$ in an ion trap system at any point in time $t$ is thus given by:
\begin{align}
    \label{eq:total_system_energy}
    E_{\mathrm{tot}}=&\sum_i\biggl[ V_{\mathrm{DC}}(\vec{r}_i)+V_{\mathrm{rf}}(\vec{r}_i,t)+\biggr.\\ \nonumber
     \biggl. &V_{\mathrm{kin}}(v_i) + \frac{1}{2}\sum_{j\neq i}V_{\mathrm{Coul}}(\vec{r}_i,\vec{r}_j)  \biggr]
\end{align}
for ions with positions $\vec{r}_i$ and velocities $v_i$. The kinetic energy for an ion with mass $m_i$ is given by $V_{\mathrm{kin}}=(1/2) m_i v_i^2$. The Coulomb interaction energy $V_{\mathrm{Coul}}$ between particles $i$ and $j$ with charge $q_i$ and $q_j$ is given by:
\begin{equation}
    V_{\mathrm{Coul}}(\vec{r}_i,\vec{r}_j) = \frac{1}{4\pi\epsilon_0}\frac{q_i q_j}{|\vec{r}_i -\vec{r}_j|},
\end{equation}
with the vacuum permittivity $\epsilon_0$.

We calculate the energy of the simulated system $E_{\mathrm{tot}}$ by inserting the numerically obtained ion positions and velocities into Equation \ref{eq:total_system_energy}. However, the relevant quantity of energy is the secular component of $E_{\mathrm{tot}}$ in Eq. \ref{eq:energy_expansion}, as rf heating is defined by changes in this secular energy. To calculate the secular energy $E_{\mathrm{sec}}$ from the particles simulated positions and velocities, Eq. \ref{eq:total_system_energy} is adjusted in two ways:

1) The rf potential $V_{\mathrm{rf}}(\vec{r}_i,t)$ is replaced with the time-averaged field as experienced by the ions, the pseudopotential, which is an effective static potential $V_{i,\mathrm{ps}}$ given by
\begin{equation}
    \label{eq:pseudopotential}
    V_{i,\mathrm{ps}}(\vec{r_i})=\frac{q^2}{4m_i\Omega_{\mathrm{rf}}^2}\left|\nabla V_{\mathrm{rf},0}(\vec{r_i})\right|^2,
\end{equation}
for an rf drive with position-dependent rf potential amplitude $V_{\mathrm{rf},0}(\vec{r_i})$ and frequency $\Omega_{\mathrm{rf}}$. For clarity, this potential is used as a means to extract the secular energy from simulation data, and is not used in the simulation itself. 

2) The positions and velocities of the ions, $\vec{r}_i$ and $\vec{v}_i$ are replaced with their secular components $\vec{r}_i^{(\mathrm{sec})}$ and $\vec{v}_i^{(\mathrm{sec})}$. These components are found by removing the rf component from the simulated positions $\vec{r}_i^{(0)}$, as follows: If the secular frequency $\omega_{\{x,y\}}$ is much lower than the rf drive frequency $\Omega_{\mathrm{rf}}$, the rf component of position $\vec{r}_i^{(\mathrm{rf})}$ can be described by the equation of motion
\begin{equation}
    \frac{d^2 \vec{r}_i^{(\mathrm{rf})}}{dt^2} \approx -\frac{q\nabla V_{\mathrm{rf},0}(\vec{r}_i^{(\mathrm{sec})})}{m}\cos{(\Omega_{\mathrm{rf}}t)}.
\end{equation}
This approximation is valid if the amplitude of $\vec{r}_i^{(\mathrm{rf})}$ in one oscillation period is small enough such that $\nabla V_{\mathrm{rf},0}(\vec{r}_i^{(\mathrm{sec})})$ is approximately constant. Simulations do not directly provide $\vec{r}_i^{(\mathrm{sec})}$, so we use an iterative approach and initially use the simulated positions $\vec{r}_i^{(0)}$ as an approximation for the secular motion: $\vec{r}_i^{(\mathrm{sec})}\approx\vec{r}_i^{(0)}$. The rf component of the position is then
\begin{equation}
    \label{eq:rf_motion_estimate}
    \vec{r}_i^{(\mathrm{rf})} \approx -\frac{q\nabla V_{\mathrm{rf},0}(\vec{r}_i^{(0)})}{m\Omega_{\mathrm{rf}}^2}\cos(\Omega_{\mathrm{rf}}t).
\end{equation}
The secular motion is iteratively approximated by removing the rf component from the full simulated positions:
\begin{align}
    \label{eq:rf_removal_first}
   \vec{r}_i^{(1)}&=\vec{r}_i^{(0)}- \vec{r}_i^{(\mathrm{rf})}\\ \nonumber
   &=\vec{r}_i^{(0)} + \frac{q\nabla V_{\mathrm{rf},0}(\vec{r}_i^{(0)})}{m\Omega_{\mathrm{rf}}^2}\cos(\Omega_{\mathrm{rf}}t).
\end{align}
As $\vec{r}_i^{(1)}$ is a better approximation for secular motion than $\vec{r}_i^{(0)}$, we can improve our estimate for $\vec{r}_i^{(\mathrm{rf})}$ in Eq. \ref{eq:rf_motion_estimate}.
Higher order adjustments to the secular position can thus be found iteratively:
\begin{equation}
\label{eq:rf_removal}
   \vec{r}_i^{(n+1)}=\vec{r}_i^{(0)} + \frac{q\nabla V_{\mathrm{rf},0}(\vec{r}_i^{(n)})}{m\Omega_{\mathrm{rf}}^2}\cos(\Omega_{\mathrm{rf}}t).
\end{equation}
Note that Equation \ref{eq:rf_removal} remains an approximation for the secular motion and is not an exact solution even as $n\rightarrow\infty$. 

\begin{figure}
    \centering
    \includegraphics[width=0.9\linewidth]{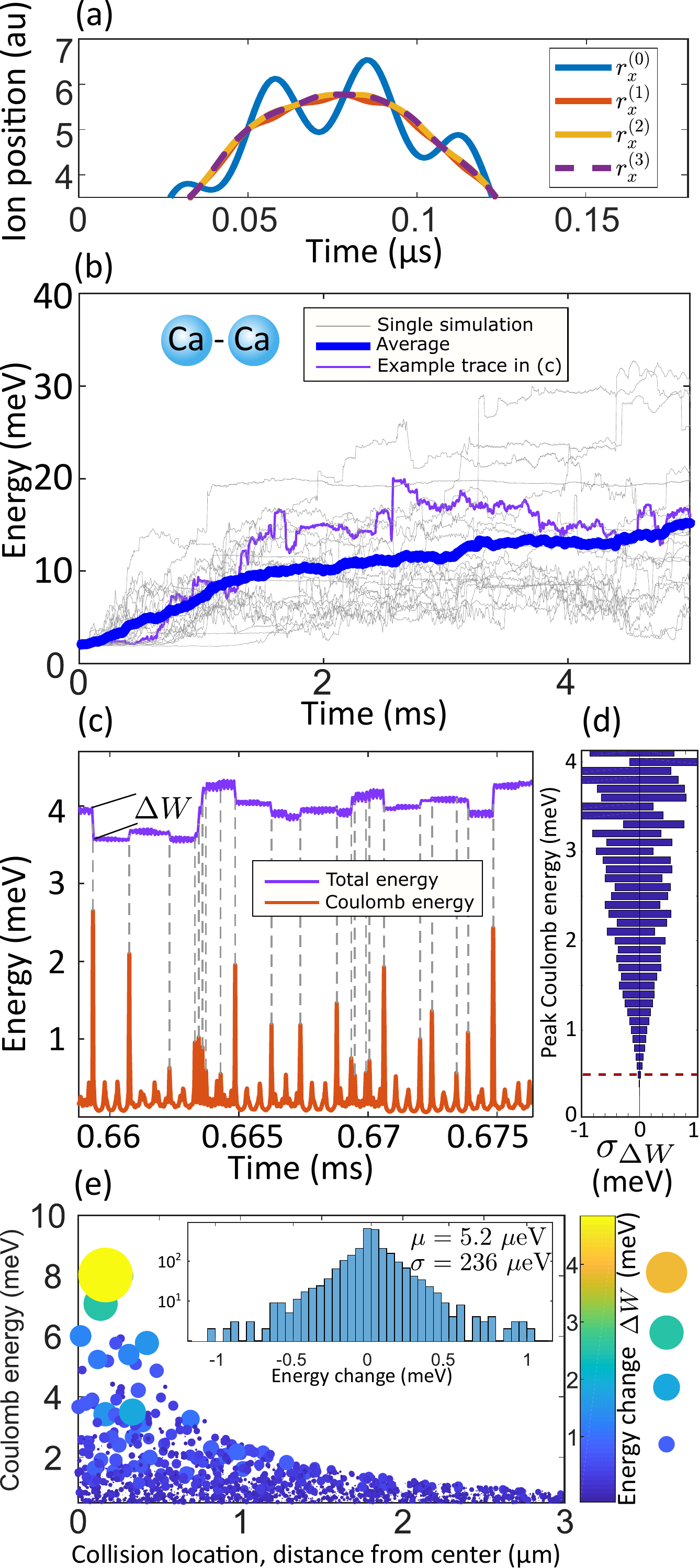}
    \caption{Analysis of energy in Ca - Ca cloud dynamics simulations. (a) To extract the non-dynamic component of energy, the rf component of the ions' motion is numerically removed. The rf-free motion is used to calculate the secular energy, $E_{\mathrm{sec}}$. (b) The development of total energy is shown (thin lines) for several simulation runs, with identical starting energies. The thick blue line is an average of the individual runs. (c) Close-up of an energy trace for one simulation, marked in (b). The discrete changes in energy ($\Delta W$) correspond to moments of high Coulomb interaction, ie. collisions. (d) Standard deviation of energy changes $\sigma_{\Delta W}$ in bins of peak Coulomb energy. (e) The Coulomb energy versus location within the trap. The marker size and color correspond to the system's energy change after the collision. The inset shows the histogram of these energy changes.}
    \label{fig:Energy_growth_trends}
\end{figure}

Figure \ref{fig:Energy_growth_trends}(a) demonstrates how a simulated trajectory (projected in one dimension) is adjusted using Eq. \ref{eq:rf_removal} in several iterations to remove the rf component of its motion. The remaining motion is approximately secular. The trajectories $\vec{r}_i^{(2)}$ and $\vec{r}_i^{(3)}$ are visibly indistinguishable in Figure \ref{fig:Energy_growth_trends}(a). We find that $\vec{r}_{i}^{n}$ changes negligibly for orders higher than $n=3$.

Using $\vec{r}_i^{(n)}\approx\vec{r}_{i}^{(\mathrm{sec})}$, the corresponding velocities $\vec{v}_i^{(n)}$, and the pseudopotential approximation $V_{i,\mathrm{ps}}$, Equation \ref{eq:total_system_energy} can be adapted to calculate $E_{\mathrm{sec}}$ for each time-step of the full simulation:
\begin{align}
    \label{eq:secular_energy}
    E_{\mathrm{sec}}=&\sum_i\biggl[ V_{\mathrm{DC}}(\vec{r}_i^{(n)})+V_{i,\mathrm{ps}}(\vec{r}_i^{(n)})+\biggr.\\ \nonumber
     \biggl. &V_{\mathrm{kin}}(v_i^{(n)}) + \frac{1}{2}\sum_{j\neq i}V_{\mathrm{Coul}}(\vec{r}_i^{(n)},\vec{r}_j^{(n)})  \biggr].
\end{align}
In this work, when describing the ion cloud's energy, we refer to the secular energy, $E_{\mathrm{sec}}$, with $n=3$.

Figure \ref{fig:Energy_growth_trends}(b) shows traces of the energy $E_{\mathrm{sec}}$ over time, for several simulation runs. All simulations start with identical parameters, except for a randomly chosen rf phase, which reflects that a collision with a background particle can occur at any time during the rf-drive cycle. In every trace, energy increases over time, but not necessarily continuously. Although the only difference between the individual simulations is the initial rf phase, there is a large variation in the development of energy over time, resulting in energies ranging from about 4 to 30 meV after 5 ms. This variation attests the chaotic nature of melted ion dynamics.

The thick blue line is an average of the individual simulations, which increases approximately with the square-root of time. To further examine the dynamics that lead to energy changes, we take a trace from Figure \ref{fig:Energy_growth_trends}(b) as an example and investigate it in Figure \ref{fig:Energy_growth_trends}(c).

Here, one can see that the secular energy does not change continuously, but at discrete points in time, corresponding to moments of increased Coulomb interaction. This behavior is ubiquitous over the full simulation duration of all traces, not just the example presented in Figure \ref{fig:Energy_growth_trends}(c). We refer to these moments of increased Coulomb interaction as \emph{collisions}, since the interaction between the charged particles results in an exchange of momentum and energy. As described in Section \ref{sec:heating_overview}, this collision is inelastic, as secular energy is not conserved. We denote the individual changes in energy as a result of collisions by $\Delta W$. 

Figure \ref{fig:Energy_growth_trends}(d) shows the standard deviation in $\Delta W$, for various bins of peak Coulomb interaction energy. All traces in (b) are used for the statistics in (d). It is apparent that higher peak Coulomb interaction allows for a greater spread in resulting secular energy change. 

We observe that collisions with Coulomb interaction energies below 0.5 meV, indicated by the dotted line in (d), do not result in noticeable changes in the system's total energy. Thus, throughout this work, we will consider an ion-ion collision to be an event where the interaction energy exceeds this threshold. 

Figure \ref{fig:Energy_growth_trends}(e) shows that collisions with higher Coulomb interaction tend to occur closer to the center of the trap, mostly within 2 $\si{\micro\meter}$ (for reference, ion oscillation amplitudes range from $\sim10 - 30~\si{\micro\meter}$). This is explained by ions having the highest possible relative velocity near the center of the trap. Larger changes in system energy require higher Coulomb interaction, and therefore occur more often in the center of the trap. The inverse is not necessarily true: high Coulomb interaction does not always result in high changes in system energy, visible in Figure \ref{fig:Energy_growth_trends}(e) by the small blue points appearing in the high Coulomb region of the plot. 

The inset in Figure \ref{fig:Energy_growth_trends}(e) shows the distribution of energy change which is qualitatively symmetric. However, the mean of the distribution $\mu = 5.2~\si{\micro}$eV and average collision rate ($\bar{f}_{\mathrm{coll}}=$ 528 collisions per millisecond), leads to an overall increase of energy of approximately 14 meV after 5 ms of rf heating.

The simulations show that rf heating is not a continuous process, but occurs at discrete moments of high Coulomb interaction energies, which we refer to as ion-ion collisions. In the following section, we model the energy change due to such a collision event, and the rate at which such events occur to build a simplified model of rf heating. 

\section{Collision model}
\label{sec:collision_model}
In this section, we provide a generalized quantitative description of the parameters involved in energy dynamics in rf heating. We use this description as a basis for a model that predicts energy changes in ion-ion collisions, and the rate at which these collisions occur. This allows us to assess the contribution of various trapping parameters to rf heating rates. The model is used to create an ion cloud energy dynamics simulation that is computationally more efficient than the full simulation used in the previous section. We limit the following discussion to a single species, two-ion cloud, but the method can easily be extended to clouds of multiple mixed-species ions.

\subsection{Model parameters}
\label{subsec:model_parameters}
Although melted ions experience an interaction energy that depends on their distance, we have determined in the previous section that below a certain threshold of Coulomb energy, changes in secular energy are negligible. We thus consider the discrete events where the Coulomb interaction surpasses the threshold that is found empirically from the full simulation. We refer to such an event as a collision. We model the energy dynamics in two steps: (1) estimate the change of energy due to a collision, and (2) estimate the collision rate. 

\textbf{(1) Collision energy:} We have established in Section \ref{sec:heating_overview} that when a collision occurs, the rf field induces a change in secular energy. We draw on results from the full ion dynamics simulation presented in Section \ref{sec:full_ion_dynamics_simulations} to derive and validate a model that describes this energy change.

The change in energy $\Delta W$ of any dynamic system of particles $i$ can be expressed in terms of the forces $\vec{F}_{i}$ acting on the particles with velocities $\vec{v}_i$, as
\begin{equation}
    \label{eq:generic_energy}
    \Delta W = \int \sum\nolimits_{i}{\vec{F}_{i}\cdot\vec{v}_{i}} dt.
\end{equation}
In a Paul trap, the total force on the ions is the sum of static and rf fields and the Coulomb force, $\vec{F}_i=\vec{F}^{(\mathrm{DC})}_i+\vec{F}^{(\mathrm{rf})}_i+\vec{F}^{(\mathrm{Coul})}_i$. The velocity of the ions can be expanded into the contributions of secular and rf motion, $\vec{v}_i=\vec{v}^{(\mathrm{sec})}_i + \vec{v}^{(\mathrm{rf})}_i$. As described in Section \ref{sec:heating_overview}, the Coulomb force enables the transfer of energy between the secular and rf motion. In the product expansion of $\vec{F}_{i}\cdot\vec{v}_{i}$ in Eq. \ref{eq:generic_energy}, this transfer is reflected by changes in the components $\int \vec{F}^{(\mathrm{Coul})}\cdot\vec{v}^{(\mathrm{rf})}dt$ and $\int \vec{F}^{(\mathrm{Coul})}\cdot\vec{v}^{(\mathrm{sec})}dt$. When evaluated over the duration of a collision, these integrals are equal and opposite in value. This value is the energy transferred due to a collision. Rf heating, which is the change in secular energy due to Coulomb interaction in an rf potential, can thus be expressed as
\begin{equation}
    \label{eq:force_int_component}
    \Delta W_{\mathrm{sec}} = -\int \sum\nolimits_{i} \vec{F}^{(\mathrm{Coul})}_{i}\cdot\vec{v}^{(\mathrm{rf})}_{i} dt.
\end{equation}
One could equally well describe the change in secular energy by the development of the integral containing $v^{(\mathrm{sec})}$. However, since secular energy is temporarily stored in $E_{\mathrm{Coul}}$ during a collision, this integral contains sharp peaks at moments of high Coulomb interaction. Equation \ref{eq:force_int_component} is thus a smoother, and therefore more intuitive, representation of $\Delta W_{\mathrm{sec}}$.

Using $F^{(\mathrm{Coul})}_{1}=-F^{(\mathrm{Coul})}_{2}$ for a two-ion collision, the change in energy reduces to
\begin{equation}
    \label{eq:force_integral}
    \Delta W_{\mathrm{sec}} = -\int\vec{F}_1^{(\mathrm{Coul})}\cdot\Delta\vec{v}^{(\mathrm{rf})} dt,
\end{equation}
with the difference in rf velocity between the two ions $\Delta \vec{v}^{(\mathrm{rf})} = \vec{v}^{(\mathrm{rf})}_{1} - \vec{v}^{(\mathrm{rf})}_{2}$.

\begin{figure*}
    \centering
    \includegraphics[width=\linewidth]{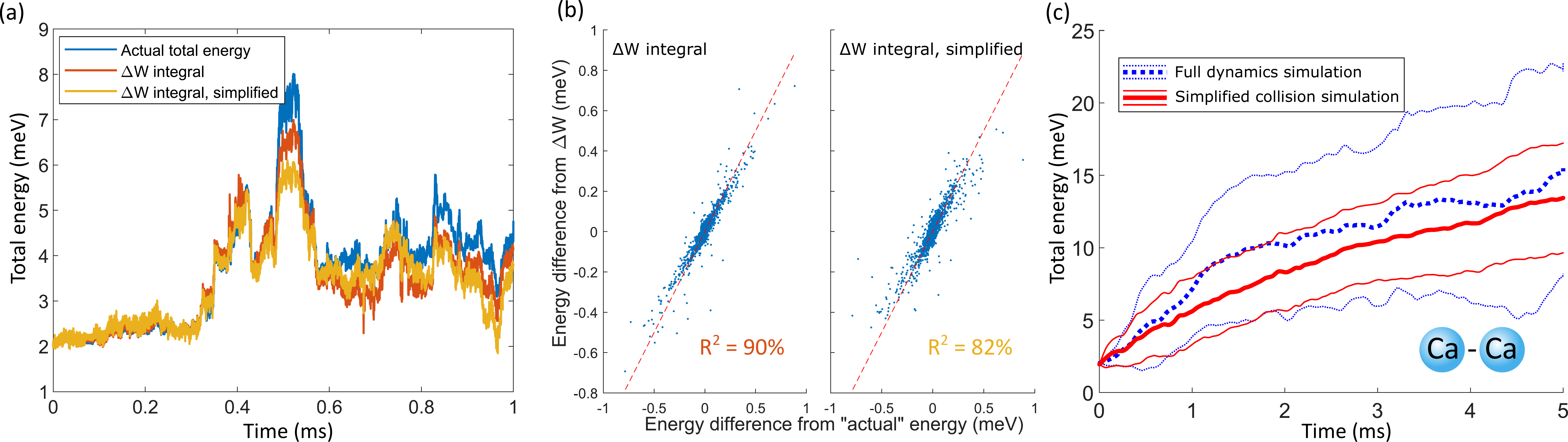}
    \caption{Comparison of energy dynamics from the simplified model and full ion dynamics simulations. (a) The total energy of an ion dynamics simulation is shown in blue. The cumulative energy from Equation \ref{eq:force_integral} is shown in red. The yellow curve uses the approximated cumulative energy Equation \ref{eq:force_integral_expanded}, with an rf-free trajectory. (b) Comparison of energy changes as determined from the calculated total energy, with those from Equations \ref{eq:force_integral} and \ref{eq:force_integral_expanded}. (c) The thick red line is the average of 20 simulation runs of the simplified energy gain model. The thin lines represent the standard deviation ($\pm\sigma$) of the individual runs. For comparison, the blue dashed curve shows the average energy from multiple full particle dynamics simulations.}
    \label{fig:Energy_force_integral}
\end{figure*}

We apply the integral of Eq. \ref{eq:force_integral} to numerical data of one of the full ion dynamics simulations from section \ref{sec:full_ion_dynamics_simulations}, which enables us to validate Eq. \ref{eq:force_integral}. Figure \ref{fig:Energy_force_integral}(a) shows the energy development as the accumulation of $\Delta W+E_{\mathrm{init}}$ (red line), with $E_{\mathrm{init}}=2$~meV to reflect the initial energy of the system. For reference, the total energy $E_{\mathrm{sec}}$ (see equation \ref{eq:secular_energy}) of the system is also plotted (blue line), showing close agreement. A comparison of the energy differences before and after collisions using the two energy metrics is shown in Figure \ref{fig:Energy_force_integral}(b) (left). The calculated correlation between the two metrics (R-squared \cite{Freedman_2007} of 90\%) confirms that Equation \ref{eq:force_integral} can faithfully describe the change in secular energy.

Simplification of the energy transfer integral (Eq. \ref{eq:force_integral}) can be achieved by approximating $\Delta\vec{v}^{(\mathrm{rf})}$. The relative rf velocity $\Delta\vec{v}^{(\mathrm{rf})}$ is estimated knowing the ions' positions relative to each other, and the phase of the rf field:

For a saddle-type rf potential
\begin{equation}
    \label{eq:rf_saddle_potential}
    V_{\mathrm{rf}}(\vec{r}_i,t)=\frac{1}{2}\psi_{\mathrm{rf}}(r_{i,x}^2-r_{i,y}^2)\cos{(\Omega_{\mathrm{rf}}t)}
\end{equation} with potential curvature $\psi_{\mathrm{rf}}$, the force on an ion $i$ with charge $q$ at position $\vec{r}_{i}=[r_{i,x},r_{i,y},r_{i,z}]$ is given by 
\begin{align}
\vec{F}^{(\mathrm{rf})}_{i}(\vec{r}_{i},t)&=-q\nabla V_{\mathrm{rf}}\\
&=[-r_{i,x},r_{i,y},0] q\psi_{\mathrm{rf}}\cos{(\Omega_{\mathrm{rf}}t)}.    
\end{align}
We have shown in Section \ref{sec:full_ion_dynamics_simulations} that we can approximate absolute changes in $r_{i,x}$ and $r_{i,y}$ to be constant during an oscillation cycle with frequency $\Omega_{\mathrm{rf}}$. Integrating  $\int\vec{F}^{(\mathrm{rf})}dt=m\vec{v}^{(\mathrm{rf})}$  with mass $m$ allows us to approximate the rf component of the velocity as
\begin{equation}
    \vec{v}^{(\mathrm{rf})}_{i}\approx[-r_{i,x},r_{i,y},0] \frac{q\psi_{\mathrm{rf}}}{m\Omega_{\mathrm{rf}}}\sin(\Omega_{\mathrm{rf}}t)
\end{equation} and the difference in rf velocity, 
\begin{equation}
    \Delta\vec{v}^{(\mathrm{rf})}\approx[-\Delta r_{x}^{(\mathrm{sec})},\Delta r_{y}^{(\mathrm{sec})},0] \frac{q\psi_{\mathrm{rf}}}{m\Omega_{\mathrm{rf}}}\sin(\Omega_{\mathrm{rf}}t).
\end{equation}
Here $\Delta r_{x}^{(\mathrm{sec})}$ and $\Delta r_{y}^{(\mathrm{sec})}$ are the ions' separation in their secular motion. 

The Coulomb force is given by
\begin{equation}
    F_1^{(\mathrm{Coul})}=\frac{1}{4\pi\epsilon_0} \frac{q^2}{|\Delta\vec{r}|^3} \Delta \vec{r}
\end{equation}
with $\Delta \vec{r}=\vec{r}_{1}-\vec{r}_{2}$ the ions' separation, and $\epsilon_0$ the vacuum permittivity. The Coulomb force $F_1^{(\mathrm{Coul})}$ is dominated by the secular motion of the ions, such that $\Delta\vec{r}\approx\Delta\vec{r}^{(\mathrm{sec})}$. Equation \ref{eq:force_integral} can thus be approximated as
\begin{align}
    \Delta W_{\mathrm{sec}} \approx &\frac{q^3\psi_{\mathrm{rf}}}{4\pi\epsilon_0 m\Omega_{\mathrm{rf}}} \nonumber \\ 
    \label{eq:force_integral_expanded}
    &\times \int \frac{\left(\Delta r_{x}^{(\mathrm{sec})}\right)^2-\left(\Delta r_{y}^{(\mathrm{sec})}\right)^2}{\left|\Delta \vec{r}^{(\mathrm{sec})}\right|^3}\sin(\Omega_{\mathrm{rf}}t)dt.
\end{align}

We numerically evaluate Eq. \ref{eq:force_integral_expanded} using data from the full ion dynamics simulation. Results, shown in Figure \ref{fig:Energy_force_integral}(a) (yellow), are in agreement with the results generated with Eq. \ref{eq:force_integral} (red).
The energy changes in these results are compared to energy changes derived from $E_{\mathrm{sec}}$ in Figure \ref{fig:Energy_force_integral}(b) (right). 
From the correlation of the data shown in Figure \ref{fig:Energy_force_integral}(b) (R-squared of 82\%), we conclude that Equation \ref{eq:force_integral_expanded} provides a good approximation of energy change in a collision. We thus have an expression that estimates secular energy changes due to collisions, that relies on relatively little information about the trap and ions. Notably, to estimate the energy change, neither the ions' absolute position within the trap nor the rf components of their motion is required. The expression simply contains ions' relative position during a collision, and fixed trap parameters.

\textbf{(2) Collision Rate:} To predict the rate of energy change, we must determine the frequency at which collisions occur. The collision rate is estimated by calculating how often such events occur for particles with oscillatory motion in three dimensions (and no Coulomb interaction). Our method is outlined below:

We start by analyzing the one-dimensional case for two identical particles. The goal is to find the duration $\Delta t_x$ that ions are within a chosen collision threshold $r_c$ of each other within a secular oscillation cycle. We assume that the ions are moving sinusoidally at their secular frequencies. This is an approximation for ions that experience driven motion by the rf field. This additional driven motion is, however, not dominant: ion motion from simulations and analytic approximations in Section \ref{sec:full_ion_dynamics_simulations} show that for typical trapping parameters the amplitude of the rf-driven motion is less than 10\% of the amplitude of the secular motion. Furthermore, the approximation is justified since the rf component of motion of the two ions is strongly correlated when they are within the collision range. Thus, the rf motion can be neglected when estimating the duration $\Delta t_x$. Additionally, as shown in Section \ref{sec:full_ion_dynamics_simulations}, most collisions that lead to appreciable changes in energy occur near the center of the trap, where the amplitude of the rf driven motion is minimal. 

The positions of ion 1 and 2 are sinusoidal in time, with amplitudes $a_{1,x}$ and $a_{2,x}$, and a relative phase $\phi_x$. The motion of the two ions is thus given by $r_{1,x} = a_{1,x}\sin{(\omega_x t)}$ and $r_{2,x} = a_{2,x}\sin{(\omega_x t + \phi_x)}$, with the oscillation frequency $\omega_x$. The distance $d_x=a_{d,x}\sin{(\omega_x t+\phi_{d,x})}$ between the two ions is sinusoidal, with amplitude and phase
\begin{align}
    a_{d,x} &= \sqrt{a_{1,x}^2+a_{2,x}^2-a_{1,x} a_{2,x}\cos(\phi_x)}\\
    \label{eq:phi_d} \phi_{d,x} &= \tan^{-1}\left(\frac{-a_{2,x}\sin{(\phi_x)}}{a_{1,x}-a_{2,x}\cos{(\phi_x)}}\right)
\end{align}
Using $t=1/\omega_x\left(\sin^{-1}(d_x/a_{d,x})-\phi_{d_x}\right)$, the amount of time $\Delta t_x$ that the two particles are within collision range $r_c$ is
\begin{align}
\Delta t_x &= t(d_x=r_c) - t(d_x=-r_c)\\
&=\frac{2}{\omega_x}\sin^{-1}\left(\frac{r_c}{a_{d,x}}\right) \label{eq:collision_time}
\end{align}

In one dimension, non-interacting particles will be within collision range exactly twice per secular oscillation period (or be continuously within range). The collision criterion of $|d_x|<r_c$ can thus be represented by a pulse wave (a periodic rectangular function):
\begin{equation}
\label{eq:pulse_wave}
  B_x(t)=\begin{cases}
    1, & \text{if}~(t \mod T_x)<\Delta t_x.\\
    0, & \text{otherwise}.
  \end{cases}
\end{equation}
with period $T_x=\pi/\omega_x$. If $r_c>a_{d,x}$, $\Delta t_x$ exceeds $T_x$, which is unphysical. Analytically, this results in $B_x$ being continuously 1, which is physically accurate.

This model can be extended to three dimensions with three pulse waves $B_{\lbrace x,y,z\rbrace}(t)$, characterized by periods $T_{\lbrace x,y,z\rbrace}$ and pulse durations $\Delta t_{\lbrace x,y,z\rbrace}$. The collision rate is described by the average pulse rate of the product of the 3 pulse waves $B_x B_y B_z$, and is given by
\begin{equation}
    \label{eq:collision_rate}
    \bar{f}_{\mathrm{coll}} = \frac{\Delta t_{x} \Delta t_{y} \Delta t_{z}}{T_x T_y T_z}\left(\frac{1}{\Delta t_{x}}+\frac{1}{\Delta t_{y}}+\frac{1}{\Delta t_{z}}\right)
\end{equation}
and average time between collisions $\bar{t}_{\mathrm{coll}} = 1/\bar{f}_{\mathrm{coll}}$. (See Appendix \ref{app:3D_collision_rate})

The condition that the three functions have an overlap ensures that the positions of the two ions are both within the bounds of a cube of sidelength $r_c$, but not necessarily that the two ions are within a distance of $r_c$ of each other. This is taken into account in our rf heating rate model, discussed in the following section.

\subsection{Simplified rf heating model}
\label{subsec:simplified_simulation}
In the previous section, we described the energy dynamics of melted ions by deriving models for ion-ion collision rates and associated energy changes. In this Section, we use these models to construct a simplified rf heating simulation in which we repeatedly generate a time until a collision occurs, $t_{\mathrm{coll}}$ (Eq. \ref{eq:collision_rate}), followed by a change in energy due to that collision, $\Delta W_{\mathrm{sec}}$ (Eq. \ref{eq:force_integral_expanded}). The simplified simulation allows us to generalize our investigation of rf heating without the computational overhead inherent in the full ion dynamics simulation. The simplified simulation is outlined below.

As in the previous section, we describe the rf-free motion of the ions $i$ as sinusoidal in three orthogonal directions (indexed $k$), with parameters $a_{1,k}$ and $a_{2,k}$ the amplitudes of ions 1 and 2, and relative phases $\phi_k$. The energy of the system, conserved as long as a collision doesn't occur, is 
\begin{equation}
    \label{eq:energy_distribution}
    E_{sec}\approx\sum_{i,k}\frac{1}{2}m\omega_k^2 a_{i,k}^2.
\end{equation}
The approximation is based on the assumption that the Coulomb energy is negligible while ions are far outside of the collision threshold. 
The simulation is initialized with a chosen energy $E_0$ distributed randomly over the amplitudes $a_{i,k}$. The initial phases $\phi_k$ are chosen randomly and uniformly. The parameters $a_{i,k}$ and $\phi_k$ characterize the state of the ions between collision events.

Applying these parameters to Equations \ref{eq:collision_time} and \ref{eq:collision_rate} yields a collision rate $\bar{f}_{\mathrm{coll}}$. The collision threshold is chosen to be $r_c=1.44$ $\si{\micro\meter}$, corresponding to a Coulomb energy of 0.5 meV, based on analysis of simulations described in section \ref{sec:full_ion_dynamics_simulations}. There we argue that Coulomb energies less than 0.5 meV do not lead to appreciable rf heating. 

The time between collisions is usually longer than a typical secular motion period and can thus be assumed to be uncorrelated due to the aperiodic nature of the motion. Therefore, the probability distribution for collision times is an exponential distribution, $P(t) =\bar{f}_{\mathrm{coll}}\exp{(-t\bar{f}_{\mathrm{coll}})}$. A random number from a known probability distribution function $P(t)$ can be generated by drawing a random value $p$, uniformly between 0 and 1, and transforming it with the inverse cumulative distribution function (ICDF) \cite{DeVroye_1986} of $P(t)$. A random time $t_{\mathrm{coll}}$ is thus generated from the ICDF of the exponential distribution, given by $-\ln{(1-p)}/\bar{f}_{\mathrm{coll}}$.

Since subsequent collisions require a pause time of at least half an oscillation period, we do not consider generated collision times lower than this period. Therefore, if the chosen  $t_{\mathrm{coll}}$ is less than $\min_k{T_k}$, a new collision time is randomly generated. The ions' oscillation amplitudes $a_{i,k}$ and phases $\phi_k$ remain unchanged until time $t_{\mathrm{coll}}$, at which a collision occurs and they need to be updated to reflect a change in energy.

We update $a_{i,k}$ and $\phi_k$ by running a Coulomb collision simulation which generates a randomized collision trajectory, based on values derived from $a_{i,k}$ and $\phi_k$. The results of the simulation, along with a randomized rf phase, are applied to Eq. \ref{eq:force_integral_expanded} to generate an energy change $\Delta W$. Based on the simulated trajectories and $\Delta W$, a new set of parameters $a_{i,k}$ and $\phi_k$ is obtained.

The Coulomb collision simulation acts as follows: For a given set $a_{i,k}$ and $\phi_k$, we determine the approximate velocity of the ions at the moment of impact. The secular velocity $\vec{v}_i=[v_x,v_y,v_z]$ of ion $i$ is given by $v_{i,k}\approx a_{i,k}\omega_{k}\cos\theta_{i,k}$, assuming that the rf contribution to the velocity is negligible for modeling a collision event.  $\theta_{i,k}$ is given by $\omega_k t$ and $\omega_k t + \phi_k$ for the two ions. Referring to Eq. \ref{eq:phi_d} in section \ref{subsec:model_parameters}, the separation between ions is given by $d_k=a_{d,k}\sin(\omega_k t+\phi_{d,k})$. During a collision, the ion separation is much smaller than the oscillation amplitude, $d_k\ll a_d$. We thus have $\omega_k t \approx -\phi_{d,k}$ during a collision (other solutions, which include integer multiples of $\pi$, can be dropped without loss of generality).  $\theta_{i,k}$ can then be written as
\begin{align}
    \theta_{1,k}&=-\tan{\left(\frac{-a_{2,k}\sin\phi_k}{a_{1,k}-a_{2,k}\cos\phi_k}\right)}\\
    \theta_{2,k}&=\theta_{1,k}+\phi_k,
\end{align}
from which we calculate $\vec{v}_i$. 

We use the ions' velocities $\vec{v}_i$ as parameters for the collision simulation, a numerical integrator in which the only force is the Coulomb interaction. Two particles are placed at random points in a box with sidelengths $r_c$, denoting their positions as $\vec{\chi}_{i}^{(0)}$. Particles are taken out of collision range by moving them to positions $\vec{\chi}_{i}^{(\mathrm{start})}=\vec{\chi}_{i}^{(0)}-\vec{v}_{i}t_s$. The time $t_s$ is chosen to be $t_s=8r_c/\mathrm{max}_{i,k}|v_{i,k}|$, where the value $8$ is chosen so that ions are placed far enough from each other such that the Coulomb energy is far below the collision threshold at start of the simulation. $\vec{\chi}_{i}^{(\mathrm{start})}$ and $\vec{v}_{i}$ are starting parameters for the simulation. The simulation is carried out for a time $2t_s$, which provides the time-dependent positions $\vec{\chi}_{i}(t)$ of the ions as they collide. A change in energy $\Delta W$ is then calculated using Eq. \ref{eq:force_integral_expanded}. Since the collision time is uncorrelated with the phase of the rf field, we add a random phase to the argument of the sine.

The calculated change in energy $\Delta W$ is used to update the parameters $a_{i,k}$ and $\phi_k$: The final velocities from the Coulomb collision simulation $v^{\mathrm{fin}}_{i,k}$ are adjusted according to Eq. \ref{eq:force_integral_expanded}, $\Delta W$. Expanding Eq. \ref{eq:force_integral_expanded} into its two sum components, the terms of containing $\Delta r_{x}^{(\mathrm{sec})}$ and $\Delta r_{y}^{(\mathrm{sec})}$ are used to adjust the values of $v^{\mathrm{fin}}_{i,x}$ and $v^{\mathrm{fin}}_{i,y}$. These adjusted velocities, together with the positions of the collision $\chi_{i,k}$, are used to calculate a new set of $a_{i,k}$ and $\phi_i$. With this updated set of parameters, a new collision time $t_{\mathrm{coll}}$ is generated. This process is repeated until the sum of all collision times exceeds the desired simulation duration. $E_{\mathrm{sec}}$ is calculated with Eq. \ref{eq:energy_distribution}, using the parameters $a_{i,k}$ from every step of the simulation, giving a time-dependent energy. The simplified simulation reduces the computation duration by more than three orders of magnitude, compared to the full ion dynamics simulation.

This method of estimating the ion cloud energy readily expands towards more than two ions by extending the parameter set $a_{i,k}$ and $\phi_{i,k}$. In this case, a collision time $t_{\mathrm{coll}}^{(i,j)}$ is generated for all combinations of ion pairs $i\neq j$, and the ion pair with the shortest collision time is selected to undergo a simulated collision. The parameters $a_{i,k}$ of the chosen ion pair, with phase difference $\phi_{i,k}-\phi_{j,k}$, are updated to reflect a collision between those two ions, using the method described above. This method is applicable if one assumes that collisions are predominantly between no more than two ions. We've determined from ion trajectory simulations (using typical experimental parameters) that for clouds of three, four, and five ions, about 3\%, 4\%, and 7\% of collisions involve three or more ions. While these percentages depend on trap parameters and ion energies, they serve as an indication of how often a more-than-two-body collision can be expected to occur.

We compare the performance of the full and simplified simulation, with identical trap parameters as used for Figure \ref{fig:Energy_growth_trends}(b). The results of the two types of simulation are displayed in Figure \ref{fig:Energy_force_integral}(c). The thick lines are averages of individual simulation runs. The average energy is in good agreement for the two simulations, though the simplified model underestimates the standard deviation of all simulation runs, shown by the thin lines, denoting one standard deviation. We have made similar comparisons for varying parameters such as motional frequencies and particle masses (not shown), and conclude that the simplified energy simulations work reliably as an indicator for average change in energy. 

\begin{figure*}
    \centering
    \includegraphics[width=\linewidth]{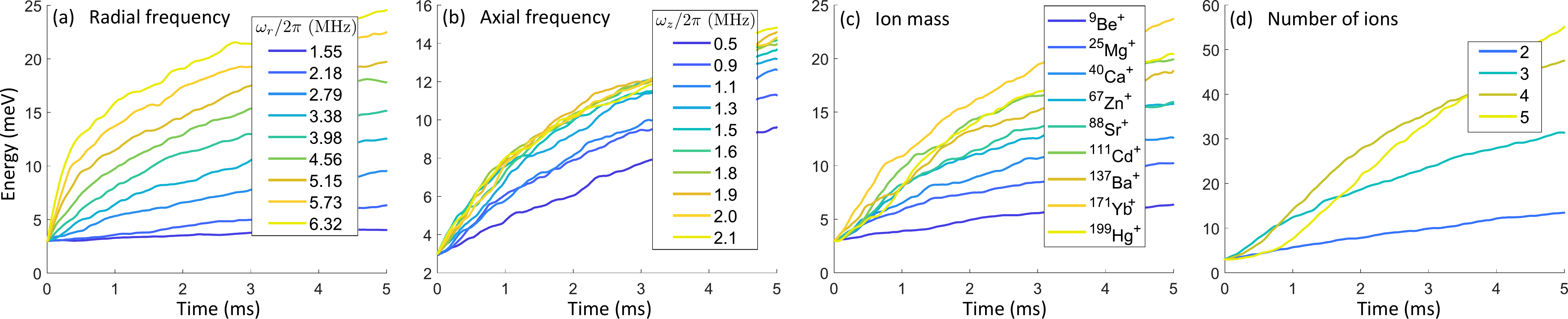}
    \caption{Energy dynamics for a two-ion cloud for various trap parameters. Unless otherwise specified, plots use the following parameters: $m=40$ amu, $\omega_z = 2\pi\cdot1.1$ MHz, $\omega_r = 2\pi\cdot3.3$ MHz, $\Omega_{\mathrm{rf}}=2\pi\cdot35$ MHz. We vary the radial motional frequencies in (a), the axial motional frequencies in (b), the ions' mass in (c) and the number of ions in (d). In (c) the trapping fields are adjusted to ensure the same motional frequencies for all masses.}
    \label{fig:MC_sims_various}
\end{figure*}

We use the simplified simulation to investigate various trapping parameters, shown in Figure \ref{fig:MC_sims_various}. Each trace is an average of 20 individual simulation runs, each with a randomly generated initial parameters $a_{i,k}$ and $\phi_k$, constrained by a fixed initial energy (3 meV), given by Equation \ref{eq:energy_distribution}. Unless otherwise noted, the simulations use two $^{40}$Ca$^{+}$ ions, with motional frequencies of $\omega_{\{x,y,z\}}=2\pi\{3.4,3.3,1.1\}$ MHz and a 35 MHz trap drive frequency. Figure \ref{fig:MC_sims_various}(a) shows traces of energy dynamics for various radial motional frequencies. Lower radial motional frequencies, and therefore a lower rf voltage, results in a lower gain in energy for a melted crystal. This behavior is observed by many ion trapping experiments, where purposefully lowering the radial confinement assists the refreezing of a melted ion crystal \cite{Hempel_PHD_2014}. As displayed in Figure \ref{fig:MC_sims_various}(b), a change in the axial confinement has a less significant influence on the rate of energy change, compared to the radial frequencies. Figure \ref{fig:MC_sims_various}(c) shows energy dynamics for various ion species, where the motional frequencies have been kept constant by adjusting the rf and DC fields accordingly. Higher masses result in higher rate of energy change. Figure \ref{fig:MC_sims_various}(d) shows the energy change for various numbers of ions. Clouds with multiple ions exhibit a larger increase in energy, as collisions are more frequent. The slower initial onset of energy increase at higher ion number is because the initial energy of 3 meV is quickly distributed over all the ions, and thus individual ions have lower average initial energies, resulting in less energetic collisions.

We generalize the results displayed in Figure \ref{fig:MC_sims_various} into a single model: We draw an analogy between melted ion energy transfer and random-walk processes, such as diffusion due to Brownian motion. In such processes, randomized changes in a variable result in an increasing statistical uncertainty in time, characterized by a diffusion constant. In our model, the energy $E_{\mathrm{sec}}$ over time $t$ follows a trend of $E\sim \sqrt{Dt}$, where $D$ is the diffusion constant \cite{Blatt_ZeitschriftFurPhysik_1986,Siemers_PhysRevA_1988}. This simple model provides an effective method to quantify the energy dynamics. To determine the parameters of the diffusive model, we perform least-squares regressions between the model $E_{\mathrm{sec}}=E_0+\sqrt{Dt}$ and our data, where $E_0$ is the initial energy (3 meV). By estimating the diffusion constant $D$ for various trap parameters, we derive a generalized expression for $D$, in terms of ion mass $m$, axial frequency $\omega_z$, radial frequency $\omega_r$, trap drive frequency $\Omega_{\mathrm{rf}}$, and number of ions $n$.

We use a polynomial model for $D$:
\begin{equation}
    D=a  m^b \omega_{r}^c \omega_{z}^d \Omega_{\mathrm{rf}}^e n^f
\end{equation}
with estimated parameters $a$ - $f$, displayed in Table \ref{tab:model_fit}.

\begin{table}[h]
\begin{tabular}{ccc}
\hline
\multicolumn{3}{c}{$E\sim\sqrt{Dt}$}                                             \\ \hline
\multicolumn{3}{c}{$D=a m^b \omega_r^c \omega_z^d \Omega_{rf}^e n^f$~(eV$^2$/s)}       \\ \hline
\multicolumn{1}{c|}{}  & \multicolumn{1}{c|}{Fit value} & Uncertainty \\
\multicolumn{1}{c|}{a} & \multicolumn{1}{c|}{330}       & 80                     \\
\multicolumn{1}{c|}{b} & \multicolumn{1}{c|}{1.0}       & 0.05                    \\
\multicolumn{1}{c|}{c} & \multicolumn{1}{c|}{2.45}      & 0.05                   \\
\multicolumn{1}{c|}{d} & \multicolumn{1}{c|}{0.52}      & 0.04                   \\
\multicolumn{1}{c|}{e} & \multicolumn{1}{c|}{0.00}       & 0.06                   \\
\multicolumn{1}{c|}{f} & \multicolumn{1}{c|}{2.96}      & 0.04                  
\end{tabular}
\caption{Fit results for rf-induced energy diffusion model}
\label{tab:model_fit}
\end{table}

For a fixed number of ions, the diffusion coefficient is most sensitive to changes in the radial motional frequency $\omega_{\mathrm{rf}}$, reinforcing the notion that reducing this parameter in an ion trap experiment (by reducing the rf voltage) is the most effective method of reducing the rf heating rate. The heating rate is to a lesser extent dependent on the axial motional frequency: this term only affects the collision rate (Eq. \ref{eq:collision_rate}), and does not influence the magnitude of energy change (Eq. \ref{eq:force_integral_expanded}). The heating rate is strongly dependent on the number of ions in the cloud, due to the increased collision rate. This conclusion emphasizes the necessity of efficient recooling when operating with increasing numbers of ions in, for example, registers of qubits for quantum computation. In the following section we investigate recooling of melted ions.

\subsection{Cooling dynamics}
\label{sec:doppler_cooling}
Typical ion trap experiments use Doppler cooling to extract energy from the ions, enabling them to become and remain crystallized in the presence of heating processes caused by electric field noise \cite{Brownnutt2015}. It is, however, often the case that the Doppler cooling parameters that cool ions close to the Doppler limit \cite{Wineland1979}, are not suitable to efficiently remove energy from an ion cloud. If the rf heating rate, described in the previous sections, is higher than the Doppler cooling rate, recrystallization will not occur.

Doppler cooling is a stochastic photon absorption and emission process, typically spanning a manifold of many electronic levels. For example, a $^{40}$Ca$^{+}$ ion is typically Doppler cooled in an eight-level manifold using 397 nm and 866 nm light. For ions with energies that are orders of magnitude higher than the Doppler cooling limit, the stochastic dynamics of this eight-level system can be approximated with a time-averaged force acting on an effective two level system. In our experiment, we typically blue-detune the 866 nm repump beam and apply an abundance of beam power. This helps avoid dark resonances \cite{Yan2019}, and allows us to approximate the eight-level system as an effective 2-level system. For a two-level system, the Doppler cooling force on an ion with velocity $\vec{v}$ is given by \cite{Metcalf_OpticsEncyclopedia_2007}:
\begin{equation}
    \label{eq:doppler_force}
    \vec{F}^{(\mathrm{D})}=\frac{\Gamma}{2}\frac{\Omega^2/2}{\Omega^2/2+\Gamma^2/4+(\delta-\vec{k}\cdot\vec{v})^2}\hbar\vec{k},
\end{equation}
with $\Gamma$ the spontaneous decay rate, $\Omega$ the on-resonance coupling strength, $\delta$ the detuning of the Doppler beam from resonance (in radians per second), $\vec{k}$ the beam's wavevector, and $\hbar$ the reduced Planck constant. 

The cooling force (Eq. \ref{eq:doppler_force}) can be included as an additional force in the full ion dynamics simulation discussed in Section \ref{sec:full_ion_dynamics_simulations}. We have compared this time-averaged force to a simulation with identical trap parameters where Doppler cooling is modeled by discrete changes in momentum due to absorption and emission of photons in a two-level system. Upon comparing the two simulations, we find that Equation \ref{eq:doppler_force} is a valid approximation of the stochastic cooling process in ion clouds. Using a time-averaged force, however, offers a lower computational cost. 

Doppler cooling dynamics can also be included in the simplified rf heating simulations. This allows for analysis of Doppler cooling parameters necessary for refreezing a melted ion chain, without the computational cost associated with full particle dynamics. Incorporating Doppler cooling in our energy gain model is not as straightforward as its inclusion in the full ion dynamics simulations, since the simplified model does not continuously track time-dependent velocities, required for calculating $\vec{F}^{(\mathrm{D})}$. We implement Doppler cooling into the simplified model as follows:

As described in Section \ref{subsec:simplified_simulation}, we determine the time between Coulomb collisions, $t_{\mathrm{coll}}$, and then make an update in particle parameters $a_{i,k}$ and $\phi_k$, based on the change in energy caused by that collision. Additional to this change, we include the change in energy due to the time-averaged Doppler cooling force 
\begin{equation}
    \label{eq:energy_change_doppler}
    \Delta W_D= \int\sum_i \vec{F}^{(\mathrm{D})}_{i}(\vec{v}_{i})\cdot\vec{v}_{i}dt.
\end{equation}
Here, we require an analytic expression for the ions' velocities $\vec{v}_{i}$. The changes in particle parameters $a_{i,k}$ due to the Doppler force are usually small between collisions, and thus they can be approximated as constant for this duration. We use this approximation to derive an analytical expression for the motion, and thus the velocities, of the ions:

An ion's motion is separable in three dimensions. Along the axial (rf-free) dimension, the ions' motion is purely secular, $r_z=a_z\sin(\omega_z t)$, and the velocity is $v_z=\omega_z a_z \cos(\omega_z t)$. Along a radial direction, the motion is composed of a secular and rf component, $r_x = r_{x,\mathrm{sec}}+r_{x,\mathrm{rf}}$, with $r_{x,\mathrm{sec}}=a_x\sin(\omega_x t)$. We find $r_{x,\mathrm{rf}}$ by reversing the approximation of Eq. \ref{eq:rf_removal_first} in Section \ref{sec:full_ion_dynamics_simulations}, which removes rf from the motion of a particle to obtain secular motion. Using $\nabla V_{\mathrm{rf},0}=\psi_{\mathrm{rf}} [r_x,-r_y,0]$ (see Eq. \ref{eq:rf_saddle_potential}), $\vec{r}^{(0)}=\vec{r}$, and $\vec{r}^{(1)}=\vec{r}^{(\mathrm{sec})}$, Eq. \ref{eq:rf_removal_first} is rewritten as:
\begin{equation}
    r_x \approx \frac{r_{x}^{(\mathrm{sec})}}{1+\frac{q\psi_{\mathrm{rf}}}{m\Omega_{\mathrm{rf}}^2}\cos{(\Omega_{\mathrm{rf}}t)}}.
\end{equation}

The velocity of ion $i$ is then given by the time-derivative of the position,

\begin{align}
    \label{eq:rf_velocity}
    v_{i,k}=&\frac{\omega_k a_{k,i}\cos{(\omega_k t)}}{1+f_k\cos{(\Omega_{\mathrm{rf}}t)}} \\ \nonumber
    +&\frac{a_{k,i}f_k\Omega_{\mathrm{rf}}\sin{(\omega_k t)}\sin{(\Omega_{\mathrm{rf}})}}{\left(1+f_k\cos{(\Omega_{\mathrm{rf}}t)}\right)^2}
\end{align}
with $f_{\{x,y\}} = q\psi_{\mathrm{rf}}/(m\Omega_{\mathrm{rf}}^2)$ and $f_z=0$. In practice, secular motional frequencies $\omega_i$ are easier to measure than the rf field curvature $\psi_{\mathrm{rf}}$, so it is useful to rewrite $f$ in terms of motional frequencies, as $f_{\{x,y\}}=\sqrt{\sum_k\omega_k^2}/\Omega_{\mathrm{rf}}$.

The analytical expression for ion velocities is used in Equation \ref{eq:energy_change_doppler} to calculate the change in energy induced by the Doppler cooling beam. Values of $a_{i,k}$ are adjusted accordingly before each collision.

\begin{figure}
    \centering
    \includegraphics[width=\linewidth]{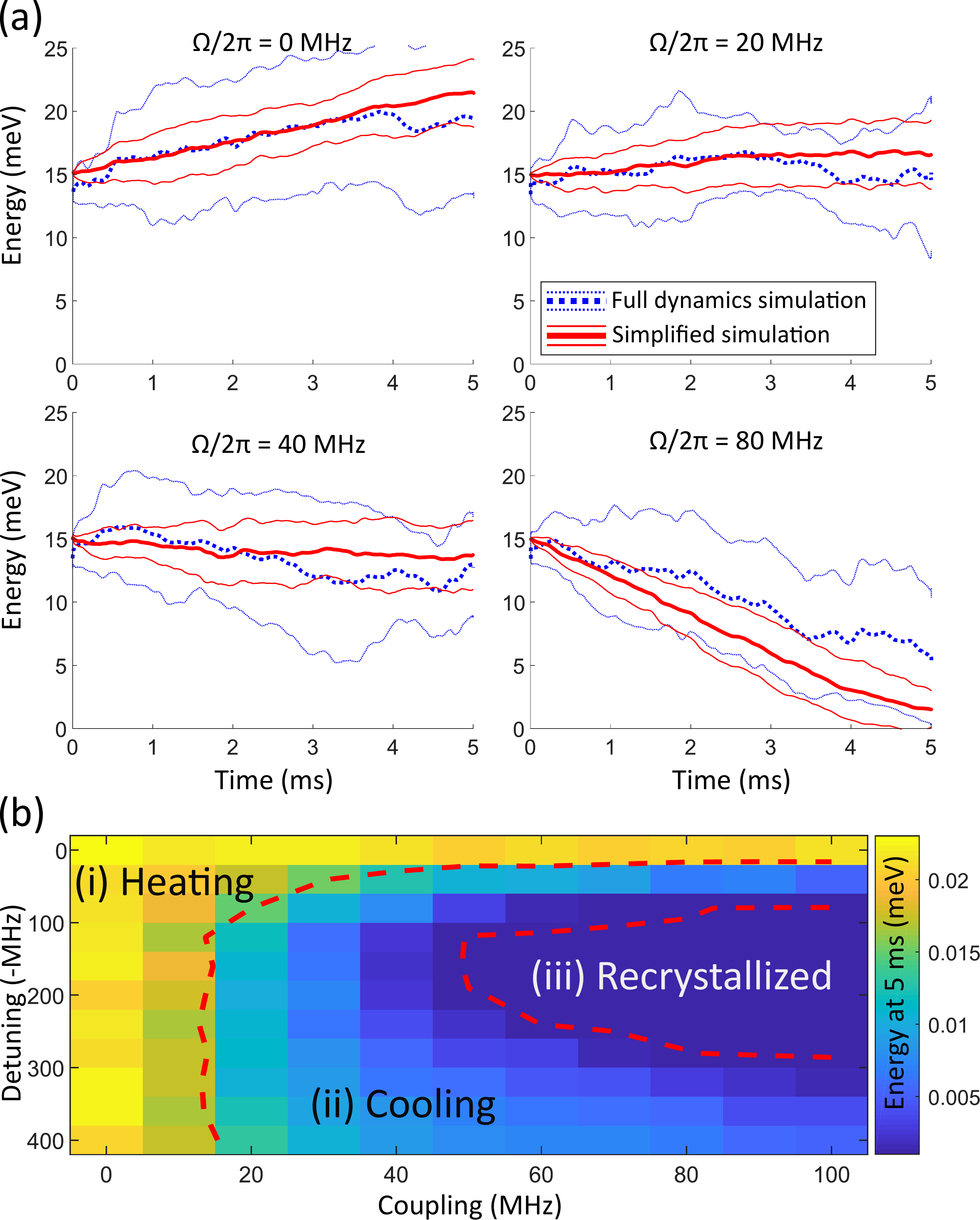}
    \caption{Simululated energy dynamics with Doppler cooling. (a) Ca - Ca cloud energy as function of time, for various Doppler cooling coupling strengths, with detuning $\delta/2\pi=-40$ MHz. For comparison, results from both the full ion dynamics simulation, and the simplified energy change simulation are displayed. The thick line is an average of multiple simulation runs, and the thin lines represents the standard deviation of the individual runs. (b) Energy after 5 ms of Doppler cooling for varying cooling strengths and detunings. For Doppler cooling parameters corresponding to regime (i), the rf heating rate is higher than the Doppler cooling rate, and recrystallization will never occur. In both (ii) and (iii), the cloud's energy is decreased, but only in (iii) are the ions recrystallized after 5 ms.}
    \label{fig:Doppler_cooling}
\end{figure}

Results of our simulations of Doppler cooling an ion cloud, with Rabi frequencies (coupling strength) $\Omega/2\pi$, ranging from 0 to 80 MHz, are shown in Figure \ref{fig:Doppler_cooling}(a), for a detuning of $\delta/2\pi=$~-40 MHz. The wavevector is chosen to be $\vec{k}=(2\pi/\lambda) [0.07,0.71,0.71]$, with $\lambda=397$~nm, which reflects the wavelength and angle of incidence in our experimental setup. We choose an initial energy of 15 meV, which is a typical ion cloud energy after 5 ms of rf heating. The thick lines in Figure \ref{fig:Doppler_cooling}(a) are averages of 20 simulation runs, and the thin lines are the standard deviation of all runs. Results from the simplified simulation are compared to results from the full dynamics simulation with identical trapping and cooling parameters (dotted lines). As in the previous section, the average trends of the two simulations are in good agreement, although the simplified model underestimates the total spread of energy in individual runs. The various plots show that with increasing coupling strength, the Doppler cooling rate overcomes the rf heating rate.

Figure \ref{fig:Doppler_cooling}(b) shows the simulated cloud (or crystal) energy after 5 ms of rf heating and simultaneous Doppler cooling, as a function of Doppler coupling strength $\Omega$ and detuning $\delta$. The initial conditions for these simulations are the same as those in Figure \ref{fig:Doppler_cooling}(a) (i.e. initial energy of 15 meV).  The figure is subdivided into three regions: in region (i) the final energy is higher than initial energy, and in regions (ii) and (iii) the final energy is lower than the initial energy. In region (iii), the final energy is low enough for the ions to recrystallize.

Although the results in Figure \ref{fig:Doppler_cooling} are specific for the chosen trap parameters, they are indicative of the magnitudes of Doppler cooling parameters required for efficient recrystallization. For example, if ions are Doppler cooled continuously, recrystallization is achieved only if $\Omega \gtrsim \Gamma \approx 2\pi$~20 MHz and $\delta \gtrsim 2\pi$~20 MHz. In typical experimental sequences such as those used for ion-based quantum computation, ions are not Doppler cooled continuously, but are separated by periods of non-cooled computation steps. If Doppler cooling is not efficient enough to recrystallize an ion cloud before these steps, the cloud subsequently reheats in the duration that the cooling beam is off. In such a cycle, ions can indefinitely remain melted. Therefore, in such sequenced experiments, the range of Doppler cooling parameters that ensure recrystallization is more stringent, corresponding to region (iii) in Figure \ref{fig:Doppler_cooling}(b). 

\section{Experimental validation}
\label{sec:experimental}
We validate the simulation methods with experimentally measured ion cloud energy dynamics due to rf heating and Doppler cooling. The goal is to demonstrate that low energy clouds undergo significant changes at the millisecond timescale, and thereby reproduce the simulated results. In our experiments, we deterministically generate ion clouds and infer their energy dynamics by monitoring their fluorescence.

Our experiments are performed on two $^{40}$Ca$^+$ ions, in a surface Paul trap (see Figure \ref{fig:paper_overview}, lower panel). Fluorescence detection and Doppler recooling is done by off-resonantly exciting the $4S_{1/2}\leftrightarrow4P_{1/2}$ transition, at 397 nm. Undesired decay from $4P_{1/2}$ to $3D_{3/2}$ is repumped with light at 866 nm. We model the cooling with an effective two-level coupling strength $\Omega$ and detuning $\delta$, calibrated using Equation \ref{eq:doppler_force} with measured fluorescence. This fluorescence is proportional to the magnitude of the Doppler cooling force of Eq. \ref{eq:doppler_force}. The power and frequency of the 397 nm light are tunable parameters, whereas the 866 nm power and wavelength are kept constant. The effective spontaneous decay rate $\Gamma$ is assumed to be dominated by the decay rate of the $4P_{1/2}\rightarrow4S_{1/2}$ transition for $^{40}\mathrm{Ca}^{+}$, and is thus $\Gamma=2\pi~21.6$~MHz. \cite{Hettrich2015}

To deterministically generate a low energy cloud, starting from a crystal, we exert a periodic force on the ions by applying an oscillating voltage on the trap electrodes, near resonance with the two radial motional frequencies, colloquially termed ``tickling'' \cite{Vedel_JournalofMassSpectrometry_1990}. We use a \textit{two}-tone signal, since ions excited in the two radial dimensions require less total energy to undergo a phase transition into a cloud, compared to excitation in one dimension. The initial cloud energy is thus lower, allowing a more accurate analysis of the cloud's energy gain. The rf tones are detuned from the motional mode frequencies by about -100 kHz. This helps to avoid recooling the crystal before it melts: Motional frequencies in our anharmonic trapping potential decrease with increasing oscillation amplitude. The oscillation frequency of ions therefore approaches resonance with the excitation field as the ions' motional energy increases.

In our experiment, we induce an oscillating electric field near the trapping region by superimposing the tickle pulse with the rf trap drive. The rf electrodes do not produce a field at the minimum of the pseudopotential, where ions are ideally located. We therefore apply a bias field of about $\nabla V_{\{x,y\}}^{(\mathrm{bias})} = 100~\si{\volt/\meter}$ in both radial directions, which displaces the ions from the trapping center, improving their coupling to the tickle field. This bias field displaces ions by $\Delta r_{\{x,y\}}=q\nabla V^{(\mathrm{bias})}_{\{x,y\}}/(m\omega_x^2)$ from the trap center, which is less than a micrometer in our setup, and is negligible when considering melted ion dynamics.

\begin{figure}
    \centering
    \includegraphics[width=\linewidth]{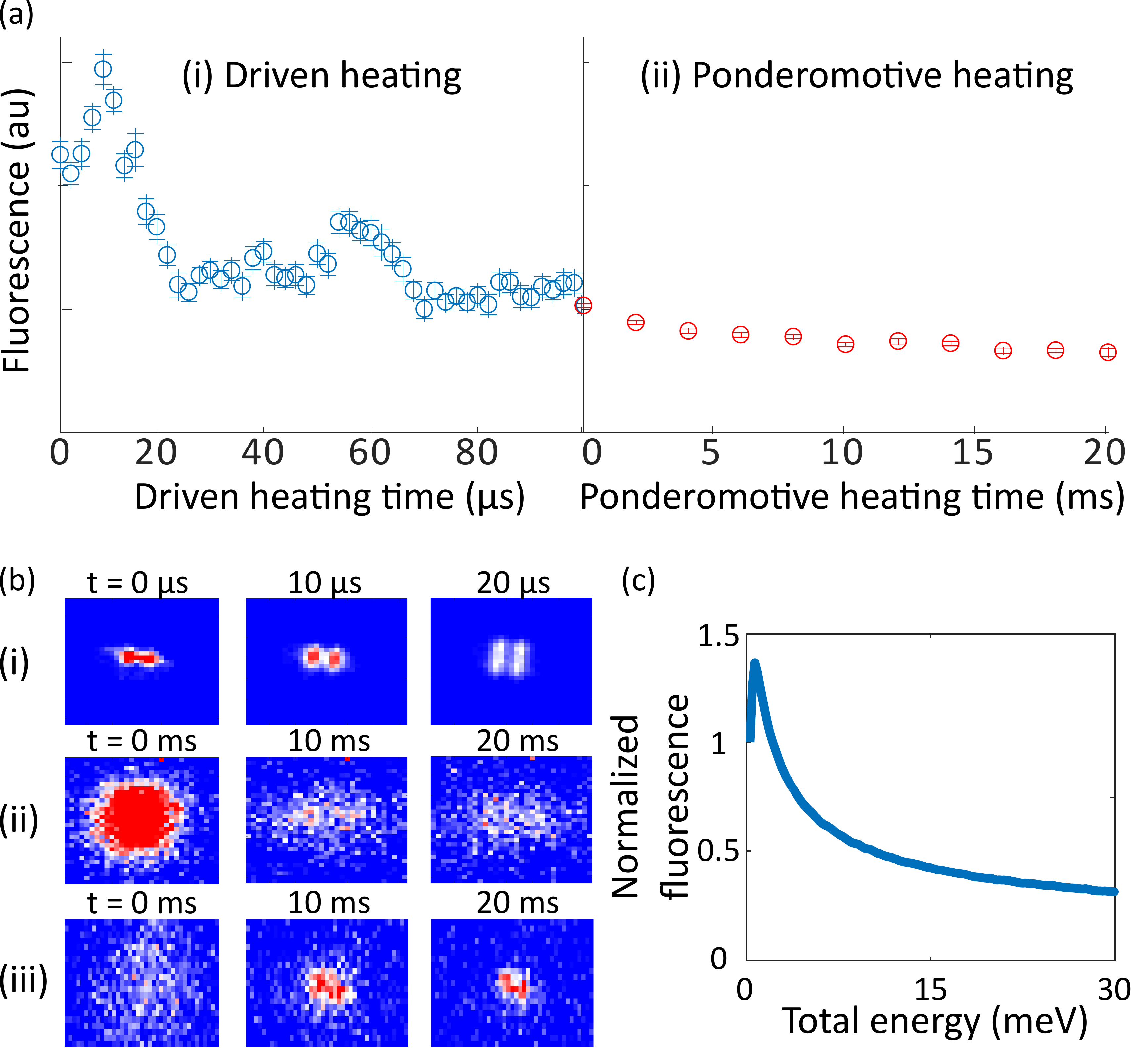}
    \caption{The Ca - Ca ion cloud energy dynamics experiments, in which ion fluorescence is monitored using (a) a PMT, and (b) a CCD camera. An experimental sequence is divided into three steps: (i) a tickle pulse excites the crystallized ions and induces enough energy to cause the crystal to melt. This is indicated by a drop in fluorescence. On the CCD camera, the two ions are no longer resolved. (ii) During an idle period, the rf field induces heating. This is detected as a drop of fluorescence and a dispersion of the cloud. (iii) Recooling the cloud with a Doppler beam, reduces the cloud size and recrystallizes the ions. (c) The fluorescence rate is simulated as function of ion energy, which allows us to interpret the measured fluorescence in terms of ion energy.}
    \label{fig:fluorescence_overview}
\end{figure}

We determine from particle dynamics simulations that roughly 5-10 meV of energy is required to melt an ion crystal when exciting it in two dimensions. In our experiment, we observe the melting by a drop in fluorescence count rate of a detection pulse, measured with a photomultiplier tube (PMT). In addition to the PMT counts, we also monitor the fluorescence on a CCD camera, providing a spatial distribution of fluorescence. The detection time must be short enough to avoid a significant change in cloud energy during the detection period. Short detection times, however, result in a lower signal-to-noise ratio in the fluorescence count rate. We choose a detection time of 500 $\si{\micro\second}$ as a trade-off. As the fluorescence rate of an ion cloud in our experiment is on the order of 10$^3$ counts per second, we detect only a few counts per sequence cycle. We thus take an average count rate from $>$1000 repetitions of the sequence.

Figure \ref{fig:fluorescence_overview}(a)(i) displays an example of detected fluorescence in terms of PMT counts, as a function of duration that the ion crystal is exposed to the tickle field. Figure \ref{fig:fluorescence_overview}(b)(i) shows corresponding CCD images (averaged over multiple shots). In these images, the horizontal axis is the axial direction, and the vertical axis is a radial direction, parallel to the trap plane. After melting, the two ions are no longer individually resolved on the CCD image.

The total energy of the ion cloud directly after melting depends on the frequencies, amplitudes and durations of both tickling pulses, and is difficult to predict exactly. Thus, in our results we do not control the initial energy of the ion cloud. However, stability in the rate of fluorescence in our measurements indicate that the average initial energy remains stable for the duration of the experiments with a fixed set of tickling signal settings.

After melting, we allow the cloud to evolve for a fixed duration, during which the ions undergo rf heating dynamics. After the evolution time, the fluorescence is probed with PMT and camera. An example of such a measurement is shown in Figure \ref{fig:fluorescence_overview}(a)(ii) and (b)(ii), where the fluorescence rate drops, and cloud size increases as function of wait time, indicating an increase in cloud energy. We subsequently apply a 20 ms pulse of a high-power ($>$100 MHz), far detuned ($\sim$120 MHz) Doppler cooling beam, which ensures that ions are recrystallized for following sequences.

The measured fluorescence count rates are to be mapped into estimated cloud energy. We find this relation through an independent simulation: a random set of oscillation amplitude parameters $a_{i,k}$ (see Section \ref{subsec:simplified_simulation}) is generated for a given ion cloud energy $E_{\mathrm{sec}}$. These amplitude parameters are used to calculate ion velocities $\vec{v}_{i,k}$ with Eq. \ref{eq:rf_velocity} for times $t\in [0,t_{\mathrm{max}}]$ with $t_{\mathrm{max}}\gg 2\pi/(\min_k \omega_k)$. Inserting these velocities into Equation \ref{eq:doppler_force} gives a time-dependent laser cooling force. The average force in the duration $t_{\mathrm{max}}$ is proportional to the fluorescence rate. Fluorescence rates are normalized to measured rates at zero energy. In the experiment, this corresponds to the rate of fluorescence detected from an unperturbed ion crystal (ie. neither displaced by a radial offset field, nor excited by means of oscillating tickle field) The procedure of simulating the fluorescence rate is repeated 20 times with random sets of $a_{i,k}$, from which we take an average.

Normalized fluorescence rates are shown in Figure \ref{fig:fluorescence_overview}(c) as a function of ion cloud energy. With this curve, a measured value of fluorescence can be used to extract the cloud energy. The mapping of fluorescence to ion energy is, however, not unique for the full domain. The measured energies, $E_{\mathrm{sec}}>5$ meV, are outside of this range of ambiguity for our parameters, $\Omega/2\pi=~64$~MHz, and $\delta/2\pi=-40$~MHz. A decrease in fluorescence rate is thus correlated with an increase in energy.

\begin{figure}
    \centering
    \includegraphics[width=\linewidth]{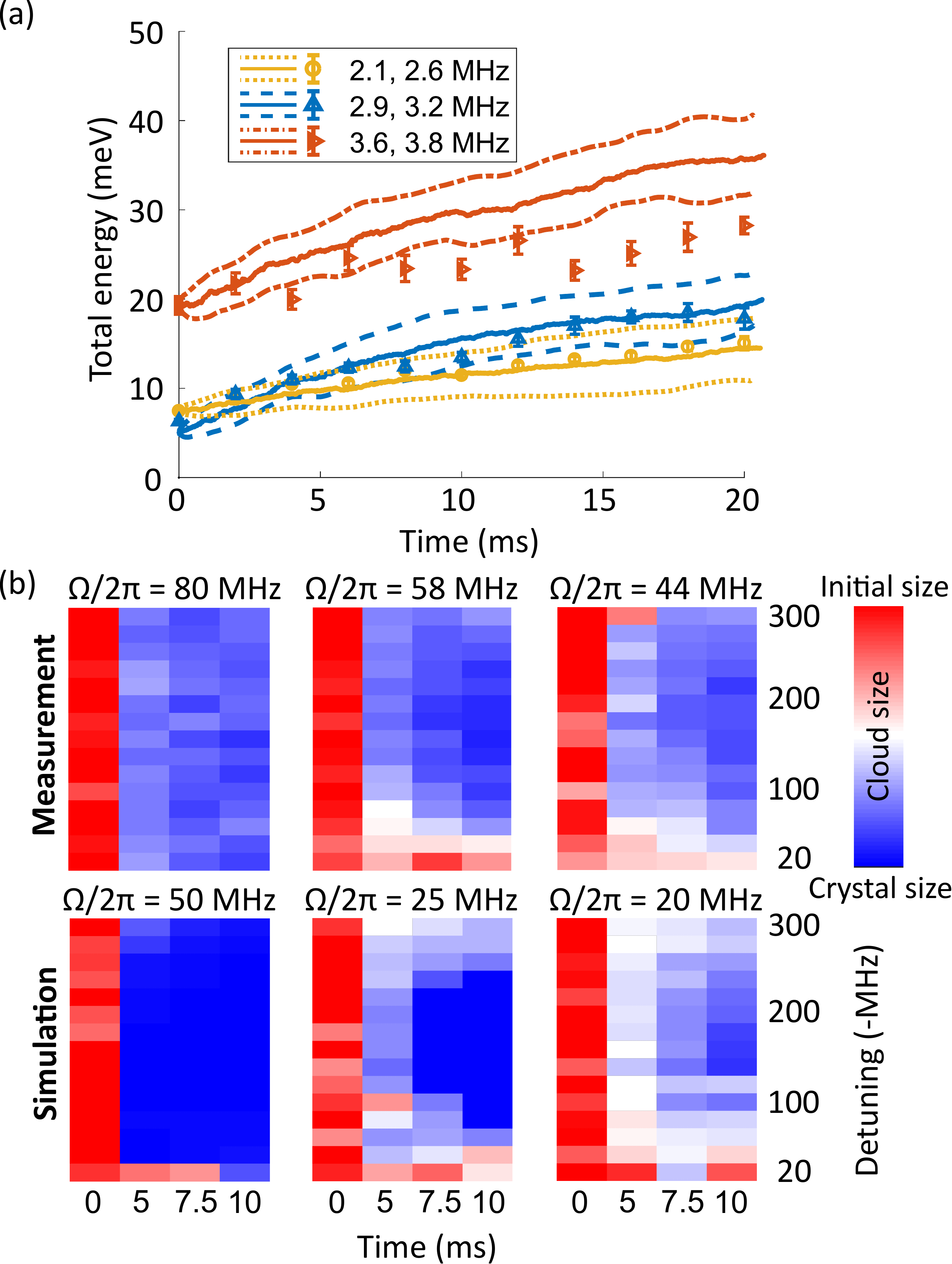}
    \caption{Ca - Ca ion cloud energy dynamics experimental results. (a) Energy dynamics have been measured for three sets of radial motional frequencies. Error bars represent the statistical spread in the acquired data, but do not take into account systematic errors in converting fluorescence into energy. The thick lines are an average of simulated energies, and the thin lines are the standard deviation ($\pm\sigma$). (b) We measure the cloud size after Doppler cooling, using CCD imaged fluorescence. We scan the Doppler beam's detuning from resonance, effective coupling strength, and Doppler cooling time. The reduction in cloud size indicates a loss of cloud energy. The colors in the figure are scaled with the mean cloud size at 0 ms and crystal size as high and low references.}
    \label{fig:fluorescence_results}
\end{figure}

Figure \ref{fig:fluorescence_results}(a) shows the ion cloud energy, inferred from measurements, as function of wait time, for various radial motional frequencies. These frequencies are adjusted by changing the power of the rf drive. The lines represent the lower and upper boundary of the standard deviation of multiple simulation runs, using the simplified rf heating model presented in Section \ref{subsec:simplified_simulation}. Measured and simulated data are in agreement for both the time-evolution of energy and the motional frequency dependence.

We apply a Doppler cooling beam after allowing the crystal to gain energy, to investigate the recooling efficiency. Variable parameters are the beam's coupling strength $\Omega$ (varied by adjusting the beam power), the beam's detuning from resonance $\delta$, and the cooling duration. As before, the energy of the cloud (or potentially crystal, after sufficient cooling) is probed with a short detection pulse. In our experiment, the detection pulse has the same coupling strength $\Omega$ and detuning $\delta$ as the cooling pulse, which means that we cannot measure the fluorescence rate independently of cooling parameters. We thus take CCD image data, and use the cloud size (using a 2D Gaussian fit) as an indicator of cloud energy. Figure \ref{fig:fluorescence_overview}(b)(iii) shows example images of an ion cloud decreasing in size with increasing recooling duration.

Figure \ref{fig:fluorescence_results}(b) displays the measured time-evolution of ion cloud size when Doppler cooling is applied. We investigate various coupling strengths $\Omega/(2\pi)=\lbrace44,~58,~80\rbrace$ MHz, at detunings ranging from $\delta/(2\pi)=$~-20 to -300 MHz. Plotted alongside are cloud sizes as determined from the simplified energy simulations (Section \ref{subsec:simplified_simulation}), including Doppler cooling. In these simulations, we reconstruct the cloud size by calculating and correlating the ions' positions and fluorescence rates using the ion motion parameters $a_{i,k}$ within the respective detection window. Cloud images are simulated at the same detunings and times as used for the experimental results shown in Figure \ref{fig:fluorescence_results}(b). The simulations are run for various coupling strength parameters, ranging from 10 to 120 MHz. From this set, we extrapolate which value of simulation coupling strength has the best agreement with the displayed experimental data, in terms of least-squares difference. The simulation results with the best agreements are displayed alongside the respective experimental results in Figure \ref{fig:fluorescence_results}(b). The color-scaling is chosen such that the two color extremes represent the average size of the initial cloud and of the crystallized ions, as detected by the CCD. We do not attribute an absolute pixel value to this scale, as the detected cloud size is dependent on cooling parameters. The perceived initial cloud size is therefore not identical for the various plots.

The simulated coupling strength values differ from the experimental values by about a factor of two. We attribute this discrepancy to two causes: 1) In our simulations, we do not correct for the spatial dependency of the power of the cooling beam, which is assumed to be uniform over the entire ion cloud. In the experiment, the beam diameter is estimated to be about 30 $\si{\micro\meter}$. At a total energy of 0.02 eV and 1 MHz axial frequency, ions typically undergo excursions of about $\pm18~\si{\micro\meter}$ from the trap center. The spatial distribution of the Doppler beam is thus not negligible. 2) The force from the Doppler cooling beam is approximated by a continuous force acting on a two-level system. In reality, this does not cover the full complexity of the stochastic forces that are described by the eight-level system. For example, in calibrating the coupling strengths $\Omega$, we fit experimental data to a model that assumes a spontaneous decay rate of $\Gamma/2\pi=21.6$~MHz, which neglects possible decay to the $3D_{3/2}$ electronic level. The modelled value of spontaneous decay $\Gamma$ is thus an upper bound for the effective two-level spontaneous decay.

Barring the discrepancy caused by the approximations used in the simulations, from both the experimental and simulated data in Figure \ref{fig:fluorescence_results}(b) the conclusion can be drawn that efficient recrystallization of a Ca - Ca cloud is achieved in about 5 ms, using Doppler cooling with $\Omega/(2\pi)>80$ MHz and $\delta/(2\pi) \approx 150$ MHz. Recrystallization is delayed or unattainable with a lower beam power and/or incorrect detuning.

\section{Conclusions}
\label{sec:conclusions}
In this work we have considered the properties of rf heating in ion clouds in Paul traps. Using a simplified simulation, we have attained a generalized model to describe the rate of energy gain after ions have melted. Experimental trials have confirmed the energy growth trends, and have confirmed the required cooling parameters for recrystallizing the ions. 

The results convey the necessity of having a recrystallization setting in experimental hardware in addition to the typical trapping and Doppler cooling settings. A detuning of half the spontaneous decay rate of the cooled ion, $\delta=\Gamma/2$ is a commonly used value for Doppler cooling in ion trap experiments, since for low coupling, $\Omega\ll\Gamma$, this detuning yields the lowest energy \cite{Wineland1979}. Figure \ref{fig:fluorescence_overview}(c) shows that this detuning is not suitable for recrystallizing an ion cloud, as the rate of rf heating exceeds the rate of Doppler cooling. It is therefore common for experiments with crystals consisting of multiple ions to have a so-called ``refreeze beam'' --- a high-power, far detuned, Doppler cooling beam. While settings for such a beam are conceptually familiar, this work provides a quantitative description of the heating and cooling processes involved.

Efficiently overcoming rf heating is imperative in ion trap systems with low ion escape barriers, such as surface traps. Ion-trap-based quantum computers envision migrating towards segmented surface trapping architectures to realize scalability \cite{Kielpinski_Nature_2002}. With increasing numbers of ions, collisions with background particles become more frequent, and therefore also the number of melting events. Even if the energy transferred in such collisions is lower than the trap depth of surface traps (typically tens or hundreds of meV), energy gain from rf heating can lead to loss of ions from the trap, possibly in tens of milliseconds. Therefore, to avoid persistently reloading ions, experimental sequences should include a refreeze phase in every cycle. Our results suggest that for recrystallization of a melted ion crystal, the Doppler cooling beam should have a detuning of roughly $\delta\approx-6\Gamma$ and should be applied for more than 5 ms, with at least $\Omega>3\Gamma$ coupling strength. Lowering the power of the rf drive field during this refreeze phase will aid recrystallization by decreasing rf heating.

These considerations become more stringent when considering mixed-species operation in surface traps, whose use is also envisioned in ion-based quantum computers \cite{Bruzewicz_npjQuantumInfo_2019}. For single-species clouds in fixed trapping fields, rf heating rates decrease as the ions' mass increases. This can be seen in Table \ref{tab:model_fit}, noting that for fixed trap parameters, $m\propto\omega_r^{-1}$. On the other hand, the trap depth, usually lowest in the radial direction perpendicular to the trap surface, is approximately proportionally lower for higher masses. Simulations show that mixed-species crystals suffer from a worst-of-both-worlds: rf heating rates are dominated by the lower mass ion in the cloud, while the trap depth remains low for the higher mass ion. This also limits the extent to which the rf drive power can be reduced without risking ion loss. It is therefore beneficial to operate ion traps with species of similar mass. However, regardless of the mixed-species mass ratio, efficient recrystallization is imperative.

In this work, we have analyzed a dynamic chaotic system and developed a simplified model to characterize it. The rf heating model can be further extended to include effects of multi-species operation, larger numbers of ions, excess micromotion, and anharmonicities in trapping potentials. Furthermore, recrystallization with Doppler cooling can be further investigated, accounting for the effective dynamics of the eight-level cooling scheme. A Doppler beam with chirped detuning \cite{Bradley_OpticsLetters_1992} can, for example, be an effective method to ensure recrystallization.

\section*{Acknowledgements}
\label{sec:acknowledgements}
This research was funded by the Office of the Director of National Intelligence (ODNI), Intelligence Advanced Research Projects Activity (IARPA), through the Army Research Office grant W911NF-16-1-0070. All statements of fact, opinions or conclusions contained herein are those of the authors and should not be construed as representing the official views or policies of IARPA, the ODNI, or the U.S. Government. We gratefully acknowledge support by the Austrian Science Fund (FWF), through the SFB BeyondC (FWF Project No. F7109-N38). We acknowledge support from the IQI GmbH.

\appendix

\section{Derivation of 3D collision rate}
\label{app:3D_collision_rate}
In Section \ref{subsec:model_parameters}, the motion of two ions is described as sinusoidal, with different amplitudes $\{a_i\}$ and frequencies $\{\omega_i\}$ in all three dimensions. In each dimension separately, ions are within collision range (such that the distance between ions $d_i$ is below a given range $r$) twice per oscillation period, $2\pi/\omega_i$. This collision condition in one dimension is thus represented by a pulse wave $B_i(t)$, with period $T_i=\pi/\omega$ and pulse duration $\Delta t_i$, as in Eq. \ref{eq:pulse_wave}. We define a two-ion collision as an event where the collision condition is satisfied in all three dimensions simultaneously, given by the pulse wave $B_{\mathrm{3D}}(t) = B_x B_y B_z$. As this three-dimensional pulse wave is aperiodic (the periods $T_i$ are not rationally related), $B_{\mathrm{3D}}(t)$ does not have a fixed pulse period. However, an \textit{average} pulse period can still be defined, given by the average time between pulses. The collision rate $\bar{f}_{\mathrm{coll}}$ is then the average number of pulses in $B_{\mathrm{3D}}$ per time. In this section, we derive the average collision rate in $B_{\mathrm{3D}}$ as function of pulse parameters $T_i$ and $\Delta t_i$. 

The probability that the 3D collision condition, $B_{\mathrm{3D}}(t)=1$, is met at any moment in time $t$ is given by the product of the probabilities that $B_i=1$ for $i={x,y,z}$, $P_{\mathrm{3D}}=\prod_i P_i$, with the 1D probabilities $P_i=\Delta t_i/T_i$. Intuitively, the collision rate is given by the product of the momentary collision probability $P_{\mathrm{3D}}$, and the effective rate at which $P_{\mathrm{3D}}$ is resampled.

\begin{figure}
    \centering
    \includegraphics[width=\linewidth]{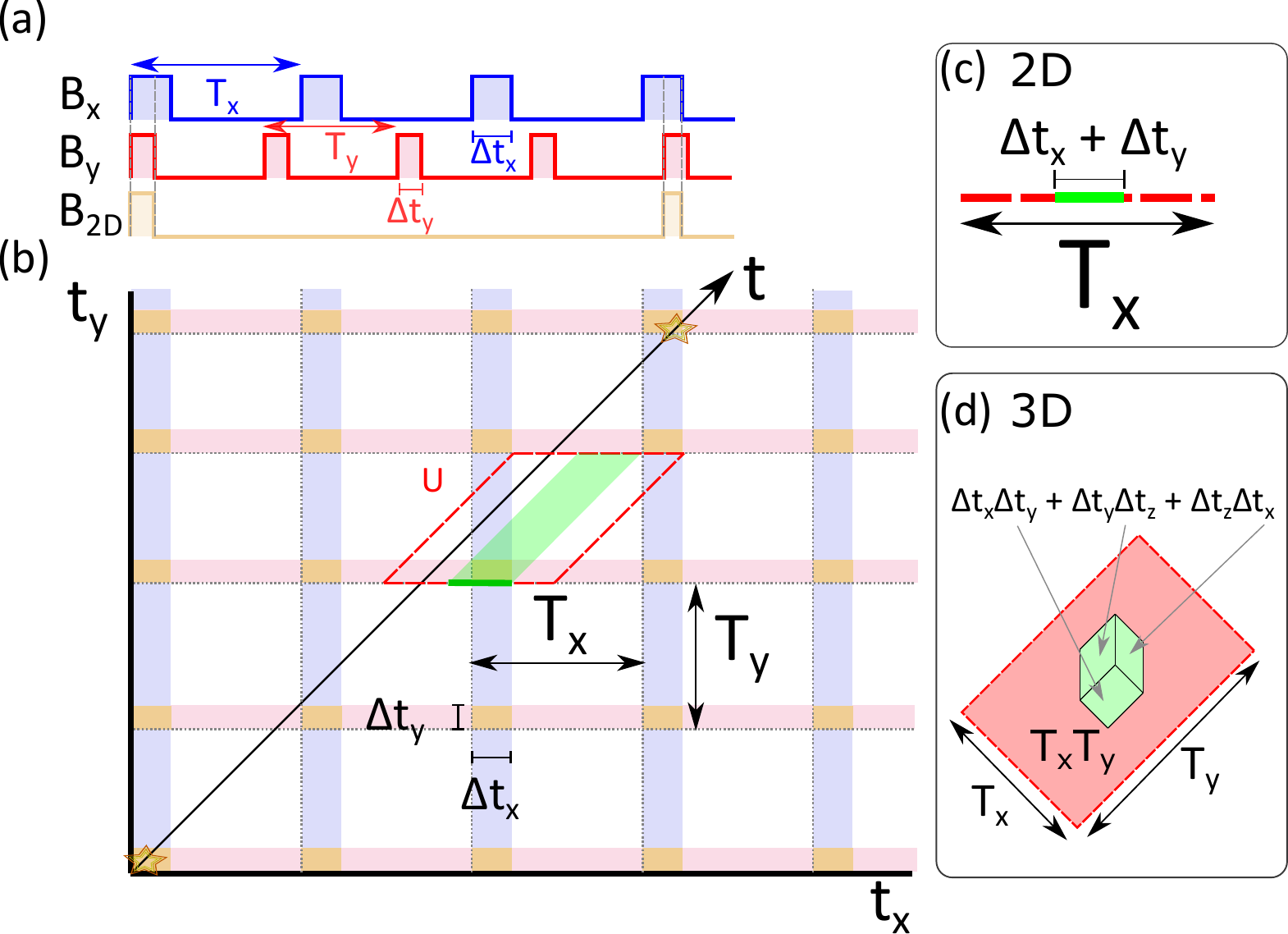}
    \caption{Graphical representation of collision condition, for deriving the 3D collision frequency. (a) In dimensions $x$ and $y$, the collision conditions are represented by pulse waves $B_x$ and $B_y$. In 2D, a collision is represented by the non-periodic pulse wave $B_{2D}=B_x B_y$. (b) $B_x(t)$ and $B_y(t)$ are displayed as two orthogonal temporal dimensions, such that their respective collision conditions are vertical and horizontal bars. ``Real'' time is the diagonal line $t_{x}=t_{y}$. Collisions occur where real time crosses vertical and horizontal bars simultaneously, denoted by the yellow boxes. This is equivalent to the real time line crossing the green shaded area, the 45$^{\circ}$ projection of collision boxes in unit cells $U$. Therefore, in the example unit cell in (b), no collision occurs. The probability of a collision occurring in a unit cell is thus given by the ratio of the projection of the green region onto the base of the unit cell, and the base of the unit cell itself. This probability, schematically shown in (c) 2D and (d) 3D, is multiplied by the frequency that the time line enters new unit cells to give the collision frequency.}
    \label{fig:collision_time_schematic}
\end{figure}

We derive $\bar{f}_{\mathrm{coll}}$ with a geometric argument, depicted in Figure \ref{fig:collision_time_schematic} (shown in two dimensions, for clarity). The pulse wave $B_{\mathrm{2D}}(t)=B_x(t)B_y(t)$ in Figure \ref{fig:collision_time_schematic}(a) is a function of time $t$. The individual pulse waves $B_x$ and $B_y$ can, however, graphically be separated into two time dimensions, $t_x$ and $t_y$, depicted in Figure \ref{fig:collision_time_schematic}(b) as two time axes. The vertical and horizontal shaded regions correspond to regions where $B_x = 1$ and $B_y = 1$, respectively. The locations where the vertical and horizontal bars meet, are places that satisfy the collision condition, $B_{\mathrm{2D}}=1$. These are graphically represented by the yellow boxes, with sidelengths $\Delta t_x$ and $\Delta t_y$. ``Real'' time $t$ parametrically follows the diagonal line, $t_x=t_y$. Whenever the real time line crosses a yellow box, a collision occurs (starred regions).

We consider the parallelogram unit cell $U$ (red dashed line), whose height is given by the lowest value of $\{T_i\}$ (in this example, the lowest value is $T_y$. We see later that this choice is made without loss of generality), and a base width given by the remaining value ($T_x$). The parallelogram angle follows the real time line $t$, at 45$^{\circ}$. Each unit cell contains exactly one collision box. The collision box is projected at 45$^{\circ}$ through the unit cell, denoted by the green shaded area. Graphically, a collision occurs if the time line passes through this area, as the real time line then has overlap with the collision box. In the example unit cell of Figure \ref{fig:collision_time_schematic}(b), no collision occurs. Since $T_x$ and $T_y$ can be assumed to be irrationally related, the location where the time line enters a unit cell is uniformly distributed. Therefore, the probability of a collision occurring in a unit cell is given by the ratio of the green shaded area to the area of the unit cell. This is identical to the ratio of the lengths of the base of the green area to the base the unit cell. The base of the green area is the projection of the collision condition box along the axis of the time line onto the base of the unit cell box (indicated by the thick green line). Since the projection is along a 45$^{\circ}$ onto the x-axis, the length of the projection is given by $\Delta t_x + \tan{(45^{\circ})}\Delta t_y=\Delta t_x + \Delta t_y$. The probability that a collision occurs within a unit cell is therefore $P_{U,\mathrm{2D}}=(\Delta t_x + \Delta t_y)/T_x$ (see Figure \ref{fig:collision_time_schematic}(c)).

Extending this concept into three dimensions (see Figure \ref{fig:collision_time_schematic}(d)), the base of the unit cell is now two dimensional, with an area of $A_U=T_x T_y$, assuming the shortest time in $\{T_i\}$ is $T_z$. The area of the projection of the collision box is given by $A_{\mathrm{coll}}=\sum_i\sum_{j>i}\Delta t_i \Delta t_j$. The probability of a collision occurring within a unit cell is $P_{U}=A_{\mathrm{coll}}/A_U$. Note that with this geometric argument $P_{U}$ can exceed 1, and should be numerically capped off at this value. For typical experimental values of $\Delta t_i$ and $T_i$ it is generally the case that $P_{U}\ll 1$.

After passing through $N$ unit cells, on average $n_{\mathrm{coll}}=P_{U}N$ collisions have occurred. The time line enters a new unit cell at intervals $T_z$, so $N=t/T_{z}$. The collision rate is thus $\bar{f}_{\mathrm{coll}}=n_{\mathrm{coll}}/t=P_{U}/T_{z}$. Rewriting gives
\begin{equation}
    \bar{f}_{\mathrm{coll}}=\prod_i \left(\frac{\Delta t_i}{T_i}\right)\sum_i\frac{1}{\Delta t_i},
\end{equation}
which conforms with the intuition that the collision rate is given by the product of $P_{\mathrm{3D}}$ and an effective resample rate.

%

\end{document}